\documentclass[lettersize,journal]{IEEEtran}
\usepackage{amsmath,amsfonts}
\usepackage{algpseudocode}
\usepackage{algorithm}
\usepackage{array}
\usepackage{textcomp}
\usepackage{stfloats}
\usepackage{url}
\usepackage{verbatim}
\usepackage{subcaption}
\usepackage{graphicx}  
\usepackage{cite}
\usepackage{float}    
\usepackage{pdfpages}
\usepackage{svg} 
\usepackage{mathrsfs}
\usepackage{verbatim}
\usepackage{amssymb}
\usepackage[justification=raggedright,singlelinecheck=false]{caption}

\begin{document}
\thispagestyle{empty}
\title{Cooperative Sensing and Communication Beamforming Design\\ for Low-Altitude Economy}

\author{Fangzhi Li, Zhichu Ren, Cunhua Pan, Hong Ren, Jing Jin, Qixing Wang, and Jiangzhou Wang

\thanks{F. Li, Z. Ren, C. Pan, H. Ren, D. Wang, and J. Wang are with National Mobile Communications Research Laboratory, Southeast University, Nanjing, China. J Jin, and Q. Wang are with the China Mobile Research Institute, Beijing, China. }
\thanks{}}

\markboth{Journal of \LaTeX\ Class Files,~Vol.~14, No.~8, August~2021}%
{Shell \MakeLowercase{\textit{et al.}}: A Sample Article Using IEEEtran.cls for IEEE Journals}


\maketitle
\thispagestyle{empty}
\begin{abstract}
To empower the low-altitude economy with high-accuracy sensing and high-rate communication, this paper proposes a cooperative integrated sensing and communication (ISAC) framework for aerial-ground networks. In the proposed system, the ground base stations (BSs) cooperatively serve the unmanned aerial vehicles (UAVs), which are equipped for either joint communication and sensing or sensing-only operations. The BSs employ coordinated beamforming to simultaneously transmit communication and sensing signals, while the UAVs execute their missions. To maximize the weighted sum rate under the sensing signal-to-interference-plus-noise ratio (SINR) constraints, we jointly optimize the transmit beamforming, receive filtering, and UAV trajectory. The resulting non-convex problem is solved using an alternating optimization framework incorporating semidefinite relaxation (SDR) and successive convex approximation (SCA). Simulation results demonstrate that the proposed joint design achieves higher communication throughput while ensuring required sensing robustness. Additionally, the sensing SINR threshold and the UAV altitude have a significant impact on the trajectory design, highlighting the necessity of adaptive deployment strategies in practical applications.
\end{abstract}


\section{Introduction}

\IEEEPARstart{T}{he} evolution of sixth-generation (6G) wireless networks is driving a paradigm shift toward architectures that unify communication and sensing functionalities. As a key enabler of this transformation, integrated sensing and communication (ISAC) leverages shared spectral and hardware resources to enable concurrent data transmission and environmental perception. This unified framework improves spectrum utilization efficiency, reduces system complexity, and enhances adaptability to environmental dynamics~\cite{zhang2024cooperative, meng2023uav, cheng2024networked}. It not only supports the convergence of sensing and communication functionalities, but also paves the way for compact and cost-efficient system designs. Consequently, ISAC has demonstrated considerable potential in various mission-critical and latency-sensitive domains, such as intelligent transportation systems, emergency response communications, and low-altitude surveillance for urban and rural monitoring applications~\cite{long2020joint}.

To further enhance ISAC system capabilities, cooperative architectures have attracted increasing attention. In contrast to single-node implementations, cooperation among multiple base stations (BSs) enables joint transmission of communication and sensing signals, facilitating coordinated beamforming, which improves spatial resolution, suppresses interference, and enhances sensing precision~\cite{zhang2024cooperative, luo2022joint}. Such cooperative schemes also enable spatial resource pooling and information sharing among BSs, thus supporting more robust and scalable ISAC deployments in heterogeneous environments. In complex scenarios with low signal-to-interference-plus-noise ratio (SINR), including urban canyons or cluttered terrains, these cooperative mechanisms significantly enhance the system's capability to detect distant or weak targets~\cite{cheng2024networked, long2020joint}. Some recent studies have further explored the integration of joint trajectory design and resource scheduling strategies for cooperative ISAC systems, offering practical modeling frameworks that serve as a foundation for algorithmic development and system-level optimization~\cite{pan2023cooperative}.

Unmanned aerial vehicles (UAVs) play an increasingly vital role in the development of cooperative ISAC systems due to their high mobility, flexible deployment, and advantageous line-of-sight (LoS) communication links~\cite{meng2023uav, jing2024isac, lyu2022joint}. UAVs can act as airborne communication relays while simultaneously engaging in sensing tasks such as target detection, environmental mapping, and trajectory tracking, making them well-suited for dynamic and infrastructure-scarce scenarios including disaster relief, environmental monitoring, and temporary deployments~\cite{khalili2024efficient, lyu2022joint, wang2024isac}. By dynamically adjusting their trajectories in response to environmental or task-specific changes, UAVs can not only enhance air-ground link quality but also extend sensing coverage, leveraging their spatial maneuverability to adapt to time-varying mission requirements and channel conditions~\cite{chai2024precoding, meng2022uav, liu2024cooperative}.

Motivated by these capabilities, cooperative ISAC systems involving multiple ground BSs and multiple heterogeneous UAVs have emerged as a promising research direction. In such systems, BSs jointly coordinate beamforming and signal scheduling to meet diverse communication demands of UAVs while concurrently enhancing sensing performance for passive targets~\cite{zhang2024cooperative, zhang2024joint, gao2024trajectory}. The heterogeneity of UAVs—in terms of capabilities, missions, and mobility—adds a layer of flexibility, enabling differentiated task allocation. However, due to the tight coupling between UAV trajectories, beamforming design, and sensing performance, the optimization of such systems becomes a highly complex, high-dimensional, and non-convex problem, especially when real-time constraints are considered~\cite{pang2024dynamic, liu2024cooperative}.

To address these challenges, researchers have proposed various joint optimization strategies combining trajectory planning and beamforming design. For example, some approaches dynamically adjust UAV positions to improve air-ground channel conditions, thereby facilitating accurate beam alignment and improved sensing results~\cite{jing2024isac, lyu2022joint, gao2024trajectory}. Others incorporate advanced physical-layer techniques, including three-dimensional beamforming, time-domain direction tracking, and target motion prediction, to further enhance both communication reliability and sensing accuracy on UAV platforms~\cite{pei2024joint, zhou2024temporal, he2023full}. Moreover, beam-assisted sensing methods have been applied to enhance UAV localization and direction-of-arrival (DoA) estimation, which are critical for maintaining robust tracking performance in dynamic or multipath-rich environments~\cite{wu2018joint}.

In the context of resource and energy optimization, several works integrate UAV energy consumption models into trajectory design, proposing energy-efficient frameworks for joint beam and trajectory control, which inform the dynamic modeling in this study~\cite{yuan2022joint}. These models allow for trade-offs between performance and endurance, especially in long-duration operations. Additionally, deep reinforcement learning (DRL) techniques have been applied to UAV-ISAC systems for adaptive control under dynamic conditions, such as moving targets, variable weather, or jamming threats. However, challenges remain regarding scalability, convergence, and interpretability~\cite{lin2023deep}. The field of beamforming is also closely linked with the development of reconfigurable intelligent surfaces (RIS). Extensive research has demonstrated that RIS can manipulate the wireless propagation environment through controllable passive reflecting elements, thereby supporting UAVs in adaptive beam steering and channel enhancement~\cite{pan2020multicell, moon2024joint, pang2021irs, cheng2024networked, deng2024joint, sankar2023beamforming, li2023joint, ge2020joint, xiu2024improving, zhang2023sensing}. By introducing additional controllable reflection paths, RIS-assisted systems expand the spatial degrees of freedom and mitigate blockage or shadowing effects in urban low-altitude scenarios. Integrating RIS into UAV-assisted ISAC systems further enables fine-grained spatial control, enhanced localization, and robust sensing performance, offering new theoretical directions and practical benefits for joint trajectory and beamforming optimization~\cite{deng2024joint, li2023joint}.

Motivated by the aforementioned challenges and opportunities, this paper presents a novel cooperative UAV-assisted ISAC system tailored for low-altitude economy (LAE) scenarios. The system incorporates two categories of UAVs—those with both communication and sensing capabilities and those designated solely for sensing tasks. Ground BSs cooperatively transmit integrated communication-sensing signals using coordinated beamforming, thereby simultaneously delivering data services and performing target detection. A unified signal model is established to characterize core processes including signal transmission, echo reception, self-interference, and receive filtering, enabling effective modeling of heterogeneous UAVs and cooperative multi-BS operations. The main contributions of this paper are summarized as follows:
\begin{itemize}
    \item We establish a cooperative UAV-assisted ISAC system model tailored for low-altitude scenarios, where multiple BSs jointly serve heterogeneous UAVs. A unified signal model is developed to capture the processes of communication transmission, echo reception, and interference, and we derive closed-form expressions for both achievable communication rate and sensing SINR.

    \item A joint optimization problem is formulated to maximize the overall communication throughput while satisfying sensing quality requirements. The optimization variables include BS transmit beamformers, UAV receive filters, and 3D UAV trajectories, subject to realistic constraints such as power limits, mobility dynamics, and collision avoidance.

    \item To efficiently solve the resulting non-convex problem, we design an alternating optimization algorithm that decomposes the problem into three submodules. Each subproblem is tackled using appropriate techniques such as semidefinite relaxation (SDR), successive convex approximation (SCA), Rayleigh quotient analysis, and trust-region methods to ensure convergence.

    \item Extensive simulations are conducted to evaluate the performance of the proposed method against baseline schemes. Results demonstrate significant gains in both communication and sensing metrics, and further sensitivity analysis offers insights into the impact of system parameters such as UAV altitude and BS density.
\end{itemize}

The structure of the paper is outlined as follows. Section II presents the system model, including the UAV architecture, communication and sensing signal modeling, and relevant constraints. Section III details the proposed solution framework, including the alternating optimization strategy and subproblem formulations for beamforming, receive filter, and trajectory design. Section IV provides simulation results that evaluate the performance of the proposed method under various system settings and compare it with several benchmark schemes. Finally, Section V concludes the paper and discusses future research directions.

\textit{Notations:} Lowercase and uppercase boldface letters represent vectors and matrices, respectively. $\mathbb{E}(\cdot)$ denotes statistical expectation. For a scalar $a$ and a vector $\mathbf{a}$, $|a|$ and $\|\mathbf{a}\|$ represent the absolute value and Euclidean norm, respectively. Superscripts $(\cdot)^{\mathrm{T}}$ and $(\cdot)^{\mathrm{H}}$ denote transpose and Hermitian transpose. $\mathbb{C}^{x \times y}$ denotes the space of $x \times y$ complex matrices. $j = \sqrt{-1}$ denotes the imaginary unit.

\section{System Model}
We consider a cooperative ISAC system, as illustrated in Fig.~1, which comprises multiple ground BSs and two types of UAVs, each serving different operational roles. The first type, referred to as communication-and-sensing UAVs, is denoted by the set $\mathcal{K}_{\mathrm{cs}} = \{1, \dots, K_{\mathrm{cs}}\}$. These UAVs are equipped with a single antenna and can perform both wireless data exchange and environmental sensing simultaneously, making them highly versatile in mission-critical scenarios. The second type, denoted as $\mathcal{K}_{\mathrm{s}} = \{1, \dots, K_{\mathrm{s}}\}$, includes sensing-only UAVs, which act as passive point targets and focus exclusively on collecting environmental information without any communication hardware. The set of all UAVs is defined as $\mathcal{K} = \mathcal{K}_{\mathrm{cs}} \cup \mathcal{K}_{\mathrm{s}}$, with a combined cardinality of $K = K_{\mathrm{cs}} + K_{\mathrm{s}}$.

The ground infrastructure consists of $M$ BSs, represented by the set $\mathcal{M} = \{1, 2, \dots, M\}$. Each BS is equipped with a uniform linear antenna array (ULA) with $L$ elements, enabling directional beamforming to support both communication and sensing. These BSs work cooperatively to serve the UAVs by transmitting integrated signals and processing returned echoes, thereby forming a tightly coupled aerial-ground ISAC network.

\begin{figure}[t] 
    \centering
    \includegraphics[width=0.4\textwidth]{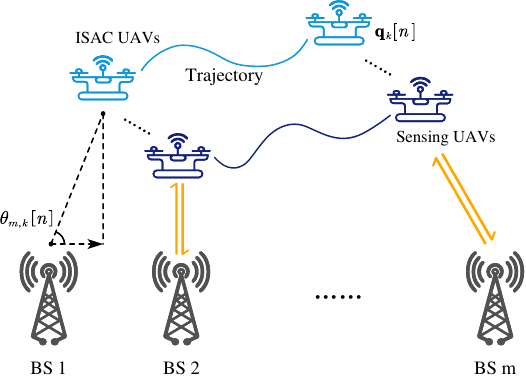} 
    \caption{System model.} 
    \label{fig:system_model} 
\end{figure}

For time management, the system operates over a total period $T$, which is divided into $N$ discrete and evenly spaced time slots. Each time slot has a duration defined as $\Delta t = \frac{T}{N}$. The collection of all these time slots is denoted by the set $\mathcal{N} = \{1, 2, \dots, N\}$, where $N$ indicates the total number of time slots within the operational period. This discretized timeline facilitates efficient modeling of dynamic interactions between the UAVs and the ground system.

In a three-dimensional Cartesian coordinate system, the horizontal position of each BS $m \in \mathcal{M}$ is represented by a two-dimensional vector $\mathbf{v}_{m} = (x_{m}, y_{m})$. Similarly, the position of the $k$-th UAV at the $n$-th time slot is characterized by its horizontal coordinates $\mathbf{q}_{k}[n] = (x_{k}[n], y_{k}[n])$ and its altitude $H_{k}$. The altitude $H_{k}$ of UAV $k$ is constrained within a specific range to ensure operational safety and compliance with system requirements:
\begin{equation}
    H_\mathrm{min} \leq H_{k} \leq H_\mathrm{max},
\end{equation}
where $H_\mathrm{min}$ and $H_\mathrm{max}$ represent the minimum and maximum permissible altitudes, respectively.

At a specific time slot $n$, the angle of departure (AoD) from BS $m$ to UAV $k$ can be computed geometrically based on the relative positions of the BS and UAV. The AoD $\theta_{m,k}[n]$ is given by
\begin{equation}
    \theta_{m,k}[n] = \arccos{\left(\frac{H_{k}}{\sqrt{\Vert \mathbf{q}_{k}[n] - \mathbf{v}_{m} \Vert^2 + H_{k}^2}}\right)}.
\end{equation}

To facilitate signal processing, the steering vector corresponding to the calculated AoD $\theta_{m,k}[n]$ is defined as
\begin{equation}
    \mathbf{a}_{m,k}[n] = [1, e^{2\pi \frac{d}{\lambda} \cos{\theta_{m,k}[n]}}, \dots, e^{2\pi \frac{d}{\lambda} (L-1) \cos{\theta_{m,k}[n]}}]^{T},
\end{equation}
where $d$ denotes the antenna spacing, and $\lambda$ is the wavelength of the transmitted signal.

Since the system operates in a low-altitude environment, non-line-of-sight (NLoS) paths are negligible. Therefore, only the LoS component is considered in the channel model, which captures the geometric relationship between the UAV and the BS n, as well as the effects of path loss and antenna steering.

The channel vector between BS $m$ and UAV $k$ at time slot $n$ is given by
\begin{equation}
    \mathbf{h}_{m,k}[n] = \sqrt{\beta_{m,k}[n]} \cdot \mathbf{a}_{m,k}[n],
\end{equation}
where $\beta_{m,k}[n]$ represents the path loss associated with the signal propagation distance between BS $m$ and UAV $k$. The path loss $\beta_{m,k}[n]$ is calculated as
\begin{equation}
    \beta_{m,k}[n] = \kappa (\Vert \mathbf{q}_{k}[n] - \mathbf{v}_{m} \Vert^2 + H_{k}^2)^{-1},
\end{equation}
where $\kappa$ is the reference path loss at a distance of 1 meter. 

By combining these expressions, the channel vector from BS $m$ to UAV $k$ can be expressed in its final form as
\begin{equation}
    \mathbf{h}_{m,k}[n] = \kappa^{\frac{1}{2}} (\Vert \mathbf{q}_{k}[n] - \mathbf{v}_{m} \Vert^2 + H_{k}^2)^{-\frac{1}{2}} \cdot \mathbf{a}_{m,k}[n].
\end{equation}

\subsection{Communication Model}

In the proposed system, the ground BSs work cooperatively to establish robust communication links with UAVs while simultaneously supporting sensing tasks. This cooperative operation enables the efficient use of spectrum and resources, ensuring that both communication and sensing requirements are fulfilled in real time. Each BS is responsible for transmitting signals that serve both communication and sensing purposes, allowing the system to adapt dynamically to the demands of the UAV network.

The total transmitted signal of BS $m$ at time slot $n$ is given by
\begin{equation}
    \mathbf{x}_m[n] = \sum_{i \in \mathcal{K}_{cs}} \mathbf{w}_{m,i}^c[n] s_{m,i}^c[n] + \sum_{i \in \mathcal{K}} \mathbf{w}_{m,i}^r[n] s_{m,i}^r[n],
    \label{eq:x_m}
\end{equation}
where $s_{m,i}^c[n]$ and $s_{m,i}^r[n]$ denote the communication and sensing symbols transmitted from BS $m$ to UAV $i$, respectively, and $\mathbf{w}_{m,i}^c[n], \mathbf{w}_{m,i}^r[n] \in \mathbb{C}^{L \times 1}$ are the corresponding beamforming vectors.

The communication signal received by UAV $k$ at time slot $n$ is a superposition of signals transmitted from all BSs, including the desired communication signal, interference from other users’ communication signals, and interference from sensing signals. This received signal can be expressed as
\begin{equation}
\begin{aligned}
\mathbf{y}_k[n] = 
&\underbrace{\sum_{m \in \mathcal{M}} \mathbf{h}_{m,k}^H[n] \mathbf{w}_{m,k}^c[n] s_{m,k}^c[n]}_{\text{desired signal}} \\
&+ \underbrace{\sum_{m \in \mathcal{M}} \sum_{\substack{i \in \mathcal{K}_{cs} \backslash k}} \mathbf{h}_{m,k}^H[n] \mathbf{w}_{m,i}^c[n] s_{m,i}^c[n]}_{\text{multi-UAV interference}} \\
&+ \underbrace{\sum_{m \in \mathcal{M}} \sum_{i \in \mathcal{K}} \mathbf{h}_{m,k}^H[n] \mathbf{w}_{m,i}^r[n] s_{m,i}^r[n]}_{\text{sensing interference}} + \mathbf{n}_k[n],
\end{aligned}
\end{equation}
where $\mathbf{n}_k[n] \sim \mathcal{CN}(0, \sigma^2\textbf{I})$ denotes additive white Gaussian noise (AWGN) at the receiver.

Based on the received signal structure, the signal-to-interference-plus-noise ratio (SINR) for UAV $k$ at time slot $n$ is defined as shown in~\eqref{eq:SINR_c}.
\begin{figure*}
    \begin{equation}
        \gamma_{k}^c[n] = 
        \frac{\displaystyle\sum_{m \in \mathcal{M}} \vert \mathbf{h}_{m,k}^H[n] \mathbf{w}_{m,k}^c[n] \vert^2}
        {\displaystyle\sum_{m \in \mathcal{M}} \displaystyle\sum_{i \in \mathcal{K}_{cs} \backslash k} \vert \mathbf{h}_{m,k}^H[n] \mathbf{w}_{m,i}^c[n] \vert^2 
        + \displaystyle\sum_{m \in \mathcal{M}} \displaystyle\sum_{i \in \mathcal{K}} \vert \mathbf{h}_{m,k}^H[n] \mathbf{w}_{m,i}^r[n] \vert^2 
        + \sigma^2}.
    \label{eq:SINR_c}
    \end{equation}
\end{figure*}

The achievable communication rate $R_{k}[n]$ for UAV $k$ during time slot $n$ is derived based on the SINR $\gamma_{k}^c[n]$, which can be expressed as
\begin{equation}
    R_{k}[n] = \log_{2}(1 + \gamma_{k}^c[n]).
\end{equation}

To evaluate the overall communication performance of the system, the sum communication rate $R[n]$ for all UAVs at time slot $n$ is defined as:
\begin{equation}
R[n]=\sum_{k\in\mathcal{K}}\omega_{k}R_{k}[n]
    =\sum_{k\in\mathcal{K}}\omega_{k}\log_{2}(1+\gamma_{k}^c[n]).
\end{equation}
where $\omega_{k}$ is a weight factor that reflects the priority or importance of UAV $k$ in the system. By appropriately choosing $\omega_{k}$, the system can prioritize certain UAVs based on their mission-critical requirements or service demands.

This communication model provides a comprehensive framework for analyzing the performance of cooperative communication and sensing systems involving multiple BSs and UAVs. It accounts for interference, noise, and the impact of beamforming on the quality of the communication links.

\subsection{Sensing Model}

In the proposed system, we consider a scenario where each UAV is sensed by the BSs using echo signal processing techniques. The sensing task involves detecting and tracking the UAVs, leveraging the reflected signals received at the BSs. The system adopts a cooperative sensing framework, wherein multiple BSs operate under a monostatic architecture and perform sensing individually via their own transmit and receive signals.

The received signal at BS $j$ at time slot $n$ can be expressed as
\begin{align}
\mathbf{y}_{r,j}[n] ={}& 
\underbrace{\sum_{l \in \mathcal{K}} \xi_{j,l} \mathbf{A}_{j,l}[n] \mathbf{x}_{j}[n]}_{\text{echo signals from UAVs}} + \underbrace{\sqrt{\zeta_{j}} \mathbf{H}_{\text{SI},j}^{H} \mathbf{x}_{j}[n]}_{\text{self-interference}} \nonumber\\
&+ \underbrace{\sum_{i \neq j, i \in \mathcal{M}} \sqrt{\alpha_{i,j}} \mathbf{G}_{i,j}^{H} \mathbf{x}_{i}[n]}_{\text{inter-BS interference}} 
+ \mathbf{n}_{r,j},
\end{align}
where $\xi_{j,l} \sim \mathcal{CN}(0, \sigma_t^2)$ denotes the radar cross-section (RCS) of UAV $l$, and $\mathbf{A}_{j,l}[n]$ is the target response matrix defined as
\begin{align}
    \mathbf{A}_{j,l}[n]
    =\sqrt{\rho_{j,l}[n]} \cdot \mathbf{a}_{j,l}[n]\mathbf{a}^{H}_{j,l}[n],
\end{align}
where $\rho_{j,l}[n]$ represents the path loss for the link between BS $j$ and UAV $l$. The coefficient $\zeta_{j}[n]$ characterizes the residual self-interference level, and the suppression coefficient $\alpha_{i,j}$ denotes the interference signal from BS $i$ to BS $j$. The matrix $\mathbf{H}_{\text{SI},j} \in \mathbb{C}^{L \times L}$ is the self-interference channel matrix of BS $j$, the matrix $\mathbf{G}_{i,j}$ represents the direct channel from BS $i$ to BS $j$, and $\mathbf{n}_{r,j} \sim \mathcal{CN}(0, \sigma_r^2 \mathbf{I}_L)$ is the additive Gaussian noise.

To isolate the reflected signals of interest, each BS applies a receive filter $\mathbf{u}_{j,k}[n] \in \mathbb{C}^{L \times 1}$. The filtered echo signal corresponding to UAV $k$, received by BS $j$, is given by

\begin{align}
\mathbf{u}_{j,k}^{H}[n]\mathbf{y}_{r,j}[n] ={}& \mathbf{u}_{j,k}^{H}[n] \sum_{l \in \mathcal{K}} \xi_{j,l} \mathbf{A}_{j,l}[n] \mathbf{x}_{j}[n] \nonumber\\
&+ \mathbf{u}_{j,k}^{H}[n] \sqrt{\zeta_{j}} \mathbf{H}_{\text{SI},j}^{H} \mathbf{x}_{j}[n] \nonumber\\
&+ \mathbf{u}_{j,k}^{H}[n] \sum_{i \neq j, i \in \mathcal{M}}\sqrt{\alpha_{i,j}} \mathbf{G}_{i,j}^{H} \mathbf{x}_{i}[n] \nonumber\\
&+ \mathbf{u}_{j,k}^{H}[n] \mathbf{n}_{r,j}.
\end{align}

The total filtered echo signal received by all BSs for UAV $k$ is represented as $\sum_{j \in \mathcal{M}} \mathbf{u}_{j,k}^{H}[n] \mathbf{y}_{r,j}[n]$. 
Based on the filtered and aggregated echo signals, the sensing SINR serves as a key metric to evaluate the quality of the sensing performance for UAV $k$ at time slot $n$. The corresponding SINR, denoted by $\gamma_{k}^r[n]$, quantifies the ratio of the power of the desired echo signals to that of the combined interference and noise. Its explicit expression is provided in~\eqref{eq:SINR_r}.

\begin{figure*}
\begin{equation}
\resizebox{\textwidth}{!}{$
\gamma_{k}^r[n] =
\frac{
    \sigma_{t}^{2} \, \mathbb{E} \left[ \displaystyle \sum_{j \in \mathcal{M}} \left| \mathbf{u}_{j,k}^{H}[n] \xi_{j,k} \mathbf{A}_{j,k}[n] \mathbf{x}_{j}[n] \right|^2 \right]
}{
    \zeta_{j} \mathbb{E} \left[ \displaystyle \sum_{j \in \mathcal{M}} \left| \mathbf{u}_{j,k}^{H}[n] \mathbf{H}_{\text{SI},j}^{H} \mathbf{x}_{j}[n] \right|^2 \right]
    +
    \alpha_{i,j} \mathbb{E} \left[ \displaystyle \sum_{j \in \mathcal{M}} \sum_{i \ne j} \left| \mathbf{u}_{j,k}^{H}[n] \mathbf{G}_{i,j}^{H} \mathbf{x}_{i}[n] \right|^2 \right]
    +
    \sigma_r^2  \displaystyle \sum_{j \in \mathcal{M}} \left\| \mathbf{u}_{j,k}[n] \right\|^2
    +
    \sigma_{t}^{2} \, \mathbb{E} \left[ \displaystyle \sum_{j \in \mathcal{M}} \displaystyle \sum_{l \ne k} \left| \mathbf{u}_{j,k}^{H}[n] \xi_{j,l} \mathbf{A}_{j,l}[n] \mathbf{x}_{j}[n] \right|^2 \right].
}
$}
\label{eq:SINR_r}
\end{equation}
\end{figure*}

\subsection{Promblem Formulation}

In the proposed ISAC system, we aim to optimize the overall performance by jointly designing the coordinated transmit beamforming vectors $\{\mathbf{w}_{m,k}^{c}[n]\}$ and $\{\mathbf{w}_{m,k}^{r}[n]\}$, the receive filter $\{\mathbf{u}_{j,k}[n]\}$, and the trajectory $\{\mathbf{q}_{k}[n]\}$ of the UAVs. The primary objective is to maximize the weighted sum rate of the UAVs while ensuring that the sensing SINR meets a predefined minimum threshold $\Gamma$ for all the UAVs at every time slot. This formulation balances the dual objectives of sensing and communication in a shared system, reflecting the ISAC paradigm.

Since the UAVs are assigned predefined initial and final positions within the total operational period \( T \), their trajectories must comply with the following position constraints:
\begin{align}
    \mathbf{q}_{k}[1] &= \mathbf{q}_{k}^{\mathrm{I}}, \label{eq:init_position}\\
    \mathbf{q}_{k}[n] &= \mathbf{q}_{k}^{\mathrm{F}}, \label{eq:final_position}
\end{align}
where \( \mathbf{q}_{k}^{\mathrm{I}} \) and \( \mathbf{q}_{k}^{\mathrm{F}} \) denote the fixed initial and final horizontal positions of UAV \( k \), respectively. These constraints ensure that each UAV begins its mission from a designated starting location and eventually reaches its assigned destination, which may be determined by mission planning, airspace coordination, or task-specific requirements.

In addition to the initial and terminal position requirements, UAV trajectories must satisfy physical motion constraints and safety requirements throughout the flight. In particular, each UAV must comply with a maximum speed limit and maintain a safe distance from other UAVs to avoid potential collisions. The constraints are specified as follows
\begin{alignat}{2}
    &\Vert\mathbf{q}_{k}[n+1]-\mathbf{q}_{k}[n]\Vert^2 
    \leq\; V_{\text{max}} \Delta t, 
    && \forall k \in \mathcal{K},\, n \in \mathcal{N}, \label{eq:speed_constraint}\\
    &\Vert\mathbf{q}_{j}[n]\!-\!\mathbf{q}_{i}[n]\Vert^2\!+\!(H_{j}\!-\!H_{i})^2 
    \!\geq\!\; D_{\text{min}}^2, 
    && \forall j, i \!\in\! \mathcal{K},\!\, j \!\neq\! i,\,\! n \!\in\! \mathcal{N}, \label{eq:collision_avoidance}
\end{alignat}
where \( V_{\text{max}} \) represents the maximum allowable speed of a UAV, and \( \Delta t \) is the fixed time duration of each slot. The parameter \( D_{\text{min}} \) denotes the minimum safe distance that must be maintained between any two UAVs at all times, taking into account their three-dimensional positions through the inclusion of altitude differences. These constraints are essential to ensure that UAV trajectories are not only dynamically feasible under motion limits, but also satisfy airspace safety requirements to prevent mid-air collisions.

Furthermore, to maintain energy efficiency and ensure practical deployment, the total transmit power of each BS must be kept within a predefined power budget $P_{\text{max}}$. Accordingly, the transmit power constraint for BS \( m \) at time slot \( n \) is given by
\begin{equation}
    \sum_{k \in \mathcal{K}_{cs}} \|\mathbf{w}_{m,k}^{c}[n]\|^{2} + \sum_{k \in \mathcal{K}} \|\mathbf{w}_{m,k}^{r}[n]\|^{2} \leq P_{\text{max}}, \quad \forall m, n. \label{eq:power_constraint}
\end{equation}
This constraint ensures that the sum power allocated to both communication and sensing signals by BS \( m \) does not exceed its maximum transmission capacity. It serves to regulate energy consumption, promote sustainable operation, and prevent signal distortion due to hardware limitations.

Given the constraints above, the joint coordinated transmit beamforming and UAV trajectory optimization problem can be formulated as follows:
\begin{subequations}
\label{eq:problem1}
\begin{align}
    (\text{P}1): & \max_{\{\mathbf{w}_{m,k}^{c}[n],\mathbf{w}_{m,k}^{r}[n],\mathbf{u}_{j,k}[n],\mathbf{q}_{k}[n]\}} \sum_{n \in \mathcal{N}} R[n] \label{eq:problem1_a}\\
    \text{s.t.} \quad 
    & \gamma_{k}^r[n] \geq \Gamma, \quad \forall k, n, \label{eq:problem1_b}\\
    & \sum_{k \in \mathcal{K}_{cs}} \|\mathbf{w}_{m,k}^c[n]\|^2 + \sum_{k \in \mathcal{K}} \|\mathbf{w}_{m,k}^r[n]\|^2 \leq P_{\max}, \quad \forall m, n, \label{eq:problem1_c}\\
    & \mathbf{q}_{k}[1] = \mathbf{q}_{k}^{\mathrm{I}}, \label{eq:problem1_d}\\
    & \mathbf{q}_{k}[n] = \mathbf{q}_{k}^{\mathrm{F}}, \label{eq:problem1_e}\\
    & \|\mathbf{q}_{k}[n+1]-\mathbf{q}_{k}[n]\|^2 \leq V_{\text{max}} \Delta t, \quad \forall k \in \mathcal{K}, n \in \mathcal{N}, \label{eq:problem1_f}\\
    & \|\mathbf{q}_{j}[n]-\mathbf{q}_{i}[n]\|^2 + (H_{j}-H_{i})^2 \geq D_{\text{min}}^2, \nonumber\\
    & \quad \forall j, i \in \mathcal{K}, j \neq i, n \in \mathcal{N}. \label{eq:problem1_g}
\end{align}
\end{subequations}
In this formulation, the objective function in~\eqref{eq:problem1_a} aims to maximize the overall communication rate $R[n]$ across all UAVs and time slots. Constraint~\eqref{eq:problem1_b} guarantees that the sensing SINR $\gamma_{k}^r[n]$ for each UAV at every time slot remains above a predefined threshold $\Gamma$, thereby ensuring reliable sensing performance. The power constraint in~\eqref{eq:problem1_c} limits the total transmit power at each BS to avoid excessive energy consumption. Constraints~\eqref{eq:problem1_d} and~\eqref{eq:problem1_e} specify the initial and final locations of the UAVs, respectively. Constraint~\eqref{eq:problem1_f} imposes a bound on the UAVs’ movement between consecutive time slots, enforcing a speed constraint based on the maximum allowed displacement. Finally, constraint~\eqref{eq:problem1_g} ensures a minimum separation distance $D_{\text{min}}$ between any pair of UAVs to prevent potential collisions during flight.

The formulated problem (P1) is a challenging non-convex optimization problem due to several intrinsic factors. The objective function and key constraints involve non-linear functions of the beamforming vectors, receive filters, and UAV trajectories, leading to a coupled optimization structure. Besides, the presence of bilinear and quadratic terms, particularly in the expressions of SINR and mobility constraints, introduces non-convexity into both the objective and feasible set. Moreover, the transmit covariance matrices are subject to rank-one constraints, which render the problem NP-hard in general. Additionally, the coupling between spatial beamforming and UAV mobility makes the optimization variables interdependent across time slots, further increasing the solution complexity. Due to these challenges, problem (P1) cannot be tackled by closed-form solutions. Therefore, in the following section, we develop an efficient alternating optimization framework that decouples the original problem into manageable subproblems and solves them iteratively using convex approximation techniques such as SDR and SCA.

\section{Proposed Solution}

In this section, we focus on solving the joint optimization problem (P1) formulated earlier. Given the complexity of jointly optimizing the coordinated transmit beamforming $\{\mathbf{w}_{m,k}^{c}[n]\}$ and $\{\mathbf{w}_{m,k}^{r}[n]\}$, the filter design $\{\mathbf{u}_{j,k}[n]\}$, and the UAV trajectory $\{\mathbf{q}_{k}[n]\}$, we employ a structured approach combining the SCA technique and alternating optimization. This approach enables us to iteratively optimize each variable while keeping the others fixed, gradually making it converge to a locally optimal solution.

\subsection{Problem Reformulation}
From the expression in~\eqref{eq:x_m}, the filtered echo signal for UAV $k$ received by BS $j$ at time slot $n$ can be decomposed as
\begin{align}
    \mathbf{u}_{j,k}^{H}[n]\mathbf{A}_{j,k}[n]\mathbf{x}_{j}[n] = &\sum_{t \in \mathcal{K}_{cs}} \mathbf{u}_{j,k}^{H}[n]\mathbf{A}_{j,k}[n]\mathbf{w}_{j,t}^c[n]s_{j,t}^c[n] \nonumber\\
    &+\sum_{t \in \mathcal{K}} \mathbf{u}_{j,k}^{H}[n]\mathbf{A}_{j,k}[n]\mathbf{w}_{j,t}^r[n]s_{j,t}^r[n].
\end{align}

Using the orthogonality and power normalization properties of the transmitted signals, i.e., $\mathbb{E}[\vert s \vert^2] = 1$, we compute the covariance matrix of the transmitted signal $\mathbf{x}_{j}[n]$ as
\begin{align}
    \mathbf{X}_{j}[n] &= \mathbb{E}[\mathbf{x}_{j}[n]\mathbf{x}_{j}^{H}[n]] \nonumber\\
    &= \sum_{t \in \mathcal{K}_{cs}} \mathbf{w}_{j,t}^{c}[n](\mathbf{w}_{j,t}^{c}[n])^{H} + \sum_{t \in \mathcal{K}} \mathbf{w}_{j,t}^{r}[n](\mathbf{w}_{j,t}^{r}[n])^{H}.
\end{align}

Thus, the numerator of the sensing SINR $\gamma_{k}^r[n]$ can be expressed as $\sigma_{t}^{2} \sum_{j \in \mathcal{M}} \mathbf{u}_{j,k}^{H}[n] \mathbf{A}_{j,k}[n] \mathbf{X}_{j}[n] \mathbf{A}_{j,k}^{H}[n] \mathbf{u}_{j,k}[n]$.

Through applying linear filtering, matrix trace properties, and substituting the signal covariance matrix \( \mathbf{X}_j[n] \), the interference and noise terms in the denominator of \( \gamma_{k}^r[n] \) can be reformulated in a compact quadratic form. As a result, the sensing SINR expression is equivalently rewritten in~\eqref{eq:SINR_re}.

\begin{figure*}
    \begin{equation}
        \gamma_{k}^r[n] = \frac{\sigma_{t}^{2} \displaystyle\sum_{j \in \mathcal{M}} \mathbf{u}_{j,k}^{H}[n] \mathbf{A}_{j,k}[n] \mathbf{X}_{j}[n] \mathbf{A}_{j,k}^{H}[n] \mathbf{u}_{j,k}[n]}
        {\displaystyle\sum_{j \in \mathcal{M}} \mathbf{u}_{j,k}^{H}[n] \left( \zeta_{j} \mathbf{H}_{\mathrm{SI},j} \mathbf{X}_{j}[n] \mathbf{H}_{\mathrm{SI},j}^{H} + \alpha_{i,j} \sum_{\substack{i \ne j, i\in \mathcal{M}}} \mathbf{G}_{i,j} \mathbf{X}_{i}[n] \mathbf{G}_{i,j}^{H} + \sigma_{r}^{2} \mathbf{I}_{L} + \sigma_{t}^{2} \sum_{\substack{l \neq k, l \in \mathcal{K}}} \mathbf{A}_{j,l}[n] \mathbf{X}_{j}[n] \mathbf{A}_{j,l}^{H}[n] \right) \mathbf{u}_{j,k}[n]}.
    \label{eq:SINR_re}
    \end{equation}
\end{figure*}

By defining \( \mathbf{H}_{m,k}[n] = \mathbf{h}_{m,k}[n] \mathbf{h}_{m,k}^{H}[n] \), and introducing the beamforming covariance matrices as \( \mathbf{W}_{m,k}^{c}[n] = \mathbf{w}_{m,k}^{c}[n] (\mathbf{w}_{j,t}^{c}[n])^{H} \) and \( \mathbf{W}_{m,k}^{r}[n] = \mathbf{w}_{m,k}^{r}[n] (\mathbf{w}_{j,t}^{r}[n])^{H} \), where \( \mathbf{W}_{m,k}^{c}[n] \succeq 0 \), \( \mathbf{W}_{m,k}^{r}[n] \succeq 0 \), and \( \text{rank}(\mathbf{W}_{m,k}^{c}[n]) = 1 \), the transmit signal covariance matrix $\mathbf{X}_{j}[n]$ can be equivalently expressed as
\begin{align}
    \mathbf{X}_{j}[n] = \sum_{t \in \mathcal{K}_{cs}} \mathbf{W}_{j,t}^{c}[n] + \sum_{t \in \mathcal{K}} \mathbf{W}_{j,t}^{r}[n],
\end{align}
where \( \mathbf{X}_{j}[n] \succeq 0 \) and each term in the summation corresponds to the contribution from either communication or sensing beamforming signals.

With these reformulations, problem (P1) can be equivalently transformed into the following problem (P2):
\begin{subequations}
\label{eq:problem2}
\begin{align}
    (\text{P}2): &\max_{\{\mathbf{W}_{m,k}^{c}[n],\mathbf{W}_{m,k}^{r}[n],\mathbf{u}_{j,k}[n],\mathbf{q}_{k}[n]\}} \sum_{n \in \mathcal{N}} \sum_{k \in \mathcal{K}} \omega_{k} \bar{R}_{k}[n], \label{eq:problem2_a}\\
    \text{s.t.} \quad 
    &\sum_{k \in \mathcal{K}_{cs}} \text{tr}(\mathbf{W}_{m,k}^{c}[n]) + \sum_{k \in \mathcal{K}} \text{tr}(\mathbf{W}_{m,k}^{r}[n]) \leq P_{\text{max}}, \, \forall m, n, \label{eq:problem2_b}\\
    &\text{rank}(\mathbf{W}_{m,k}^{c}[n]) = 1, \, \forall m, k, n, \label{eq:problem2_c}\\
    &\text{~\eqref{eq:problem1_b}, ~\eqref{eq:problem1_d}, ~\eqref{eq:problem1_e}, ~\eqref{eq:problem1_f}, ~\eqref{eq:problem1_g}}. \nonumber
\end{align}
\end{subequations}

To solve (P2), we adopt the Alternating Optimization (AO)-based algorithm, iteratively optimizing one set of variables while fixing the others.

\subsection{Transmit Beamforming Optimization}

In this section, we propose an efficient algorithm to address problem (P2), focusing on optimizing the communication and sensing beamforming matrices $\mathbf{W}_{m,k}^{c}[n]$ and $\mathbf{W}_{m,k}^{r}[n]$. The optimization is performed under fixed filter designs $\mathbf{u}_{j,k}[n]$ and UAV trajectories $\mathbf{q}_{k}[n]$. By carefully analyzing the structure of the problem, it can be seen that optimizing the beamforming is equivalent to solving the following problem (P3):
\begin{align}
    (\text{P}3): &\max_{\{\mathbf{W}_{m,k}^{c}[n],\mathbf{W}_{m,k}^{r}[n]\}} \sum_{n\in\mathcal{N}} R[n]\nonumber\\
    \text{s.t.} &\text{~\eqref{eq:problem1_b}, ~\eqref{eq:problem2_b}, ~\eqref{eq:problem2_c}.}\nonumber
\end{align}

Constraint~\eqref{eq:problem2_b}, which represents the power budget at each BS, is straightforwardly convex as it is linear in terms of the optimization variables $\mathbf{W}_{m,k}^{c}[n]$ and $\mathbf{W}_{m,k}^{r}[n]$. To verify whether constraint~\eqref{eq:problem1_b} is convex, we reformulate it as the inequality in~\eqref{eq:ineq}.
\begin{align}
     \sigma_{t}^{2}\sum_{j\in\mathcal{M}}\mathbf{u}_{j,k}^{H}[n]\mathbf{A}_{j,k}[n]\mathbf{X}_{j}[n]\mathbf{A}_{j,k}^{H}[n]\mathbf{u}_{j,k}[n] \ge \nonumber\\
     \Gamma \cdot  \{\sum_{j\in\mathcal{M}}\mathbf{u}_{j,k}^{H}[n] (\zeta_{j}\mathbf{H}_{\mathrm{SI},j}\mathbf{X}_{j}[n]\mathbf{H}_{\mathrm{SI},j}^{H} \nonumber\\
     +\sum_{i\ne j,i\in \mathcal{M}}\alpha_{i,j}\mathbf{G}_{i,j}\mathbf{X}_{i}[n]\mathbf{G}_{i,j}^{H}+\sigma_{r}^{2}\mathbf{I}_{L} \nonumber\\
     +\sigma_{t}^{2}\sum_{l\ne k,l\in\mathcal{K}}\mathbf{A}_{j,l}[n]\mathbf{X}_{j}[n]\mathbf{A}_{j,l}^{H}[n])\mathbf{u}_{j,k}[n] \}.
     \label{eq:ineq}
\end{align}

It can be observed that this constraint is linear with respect to $\mathbf{X}_{j}[n]$. Since $\mathbf{X}_{j}[n]$ is a jointly convex function of $\mathbf{W}_{m,k}^{c}[n]$ and $\mathbf{W}_{m,k}^{r}[n]$, it follows that constraint~\eqref{eq:problem1_b} is also convex with respect to these variables. 

Although constraints~\eqref{eq:problem1_b} and~\eqref{eq:problem2_b} are convex, the objective function $R[n]$ and the rank constraint~\eqref{eq:problem2_c} introduce non-convexity into problem (P3). To address this challenge, the SCA method is applied, where the non-convex components of the objective function are approximated with convex surrogates in each iteration. 

Let $\mathbf{W}_{m,k}^{c(f)}[n]$ and $\mathbf{W}_{m,k}^{r(f)}[n]$ denote the current local point at the $f$-th iteration, $f \ge 1$. During each iteration, the logarithmic terms in the objective function are linearized around the current solution. This linearization results in a convex approximation of the objective function, which can then be solved efficiently using standard convex optimization techniques. By iteratively updating the beamforming matrices and solving the resulting subproblems, the algorithm converges to a high-quality locally optimal solution. To illustrate this, the communication SINR $\gamma_{k}^{c}[n]$ can be re-expressed as shown in~\eqref{eq:SINR_ce}:
\begin{figure*}
    \begin{equation}                
    \gamma_{k}^c[n]=\frac{\displaystyle\sum_{m\in\mathcal{M}}\text{tr}(\mathbf{H}_{m,k}[n]\mathbf{W}_{m,k}^{c}[n])}{\displaystyle\sum_{m\in\mathcal{M}}\sum_{i \in \mathcal{K}_{cs} \backslash k}\text{tr}(\mathbf{H}_{m,k}[n]\mathbf{W}_{m,i}^{c}[n])+\sum_{m\in\mathcal{M}}\sum_{i\in \mathcal{K}}\text{tr}(\mathbf{H}_{m,k}[n]\mathbf{W}_{m,i}^{r}[n])+\sigma^2}.
    \label{eq:SINR_ce}
    \end{equation}
\end{figure*}

The achievable rate $R_{k}[n]$ for UAV $k$ at time slot $n$ can then be expressed as
\begin{align}
    R_{k}[n]={}&\log_{2}(1+\gamma_{k}^{c}[n]) \nonumber\\
    ={}&\log_{2}(\sum_{m \in \mathcal{M}}\sum_{i \in \mathcal{K}_{cs}}\text{tr}(\mathbf{H}_{m,k}[n]\mathbf{W}_{m,i}^{c}[n]) \nonumber\\
    &+\sum_{m \in \mathcal{M}}\sum_{i \in \mathcal{K}}\text{tr}(\mathbf{H}_{m,k}[n]\mathbf{W}_{m,i}^{r}[n])+\sigma^2) \nonumber\\
    &-\log_{2}(\sum_{m \in \mathcal{M}}\sum_{i \in \mathcal{K}_{cs} \backslash k}\text{tr}(\mathbf{H}_{m,k}[n]\mathbf{W}_{m,i}^{c}[n]) \nonumber\\
    &+\sum_{m \in \mathcal{M}}\sum_{i \in \mathcal{K}}\text{tr}(\mathbf{H}_{m,k}[n]\mathbf{W}_{m,i}^{r}[n])+\sigma^2). \nonumber\\
    \label{eq:original_rate}
\end{align}

To address the non-convexity arising from the difference of logarithmic functions in~\eqref{eq:original_rate}, we employ the first-order Taylor expansion to approximate the second logarithmic term at a feasible point \( (\mathbf{W}_{m,i}^{c(f)}[n], \mathbf{W}_{m,i}^{r(f)}[n]) \). This yields a concave lower bound for the achievable rate \( R_{k}[n] \), expressed as
\begin{align}
    R_{k}[n]
    \ge {}& \log_{2}(\sum_{m \in \mathcal{M}}\sum_{i \in \mathcal{K}_{cs}}\text{tr}(\mathbf{H}_{m,k}[n]\mathbf{W}_{m,i}^{c}[n]) \nonumber\\
    &+\sum_{m \in \mathcal{M}}\sum_{i \in \mathcal{K}}\text{tr}(\mathbf{H}_{m,k}[n]\mathbf{W}_{m,i}^{r}[n])+\sigma^2) -a_{k}^{(f)}[n] \nonumber\\
    &-\sum_{m \in \mathcal{M}}\sum_{i \in \mathcal{K}_{cs} \backslash k}\text{tr}(\mathbf{B}_{m,k}^{(f)}[n](\mathbf{W}_{m,i}^{c}[n]-\mathbf{W}_{m,i}^{c(f)}[n])) \nonumber\\
    &-\sum_{m \in \mathcal{M}}\sum_{i \in \mathcal{K}_{cs}}\text{tr}(\mathbf{B}_{m,k}^{(f)}[n](\mathbf{W}_{m,i}^{r}[n]-\mathbf{W}_{m,i}^{r(f)}[n])) \nonumber\\
    \triangleq {}& \bar{R}_{k}[n],
\end{align}
 where
\begin{align}
    a_{k}^{(f)}[n]={}&\log_{2}(\sum_{m \in \mathcal{M}}\sum_{i \in \mathcal{K}_{cs} \backslash k} \text{tr}(\mathbf{H}_{m,k}[n]\mathbf{W}_{m,i}^{c(f)}[n]) \nonumber\\
    &+\sum_{m \in \mathcal{M}}\sum_{i \in \mathcal{K}}\text{tr}(\mathbf{H}_{m,k}[n]\mathbf{W}_{m,i}^{r(f)}[n]+\sigma^2),
\end{align}
$\mathbf{B}_{m,k}^{(f)}[n]$ is defined in~\eqref{eq:B_mk}.
\begin{figure*}
    \begin{equation}                
    \mathbf{B}_{m,k}^{(f)}[n]=\frac{\log_{2}e \cdot \mathbf{H}_{m,k}[n]}{\displaystyle\sum_{m \in \mathcal{M}}\sum_{i \in \mathcal{K}_{cs} \backslash k} \text{tr}(\mathbf{H}_{m,k}[n]\mathbf{W}_{m,i}^{c(f)}[n]) +\sum_{m \in \mathcal{M}}\sum_{i \in \mathcal{K}}\text{tr}(\mathbf{H}_{m,k}[n]\mathbf{W}_{m,i}^{r(f)}[n]+\sigma^2}.
    \label{eq:B_mk}
    \end{equation}
\end{figure*}
Based on the concave lower bound \( \bar{R}_{k}[n] \), the original weighted sum rate maximization problem can be approximated by the following convex surrogate problem
\begin{align}
    (\text{P}4):&\max_{\{\mathbf{W}_{m,k}^{c}[n],\mathbf{W}_{m,k}^{r}[n]\}}\sum_{n\in\mathcal{N}} \bar R[n]\nonumber\\
    \text{s.t.} &\text{~\eqref{eq:problem1_b}, ~\eqref{eq:problem2_b}, ~\eqref{eq:problem2_c}.}\nonumber
\end{align}

However, the rank-one constraint~\eqref{eq:problem2_c} in (P4) remains non-convex. To address this, the SDR method is employed, which relaxes the rank-one constraint and transforms (P4) into a standard convex optimization problem, referred to as (P4.SDR). The relaxed problem can be solved via CVX tools, yielding solutions $\mathbf{W}_{m,k}^{c*}[n]$ and $\mathbf{W}_{m,k}^{r*}[n]$. If these solutions are not rank-one, equivalent rank-one approximations are constructed as follows:
\begin{align}
    \bar{\mathbf{w}}_{m,i}^{c}[n]={}&\frac{\mathbf{W}_{m,i}^{c*}[n]\mathbf{h}_{m,k}[n]}{\sqrt{\mathbf{h}_{m,k}^{H}[n]\mathbf{W}_{m,i}^{c*}[n]\mathbf{h}_{m,k}[n]}}, \\
    \bar{\mathbf{W}}_{m,i}^{c}[n]={}&\bar{\mathbf{w}}_{m,i}^{c}[n] \cdot (\bar{\mathbf{w}}_{m,i}^{c}[n])^{H}, \\
    \bar{\mathbf{W}}_{m,i}^{r}[n]={}& \sum_{i \in \mathcal{K}} \mathbf{W}_{m,i}^{c*}[n] + \bar{\mathbf{W}}_{m,i}^{r*}[n]-\sum_{i \in \mathcal{K}} \bar{\mathbf{W}}_{m,i}^{c}[n].
\end{align}

These solutions are rank-one and feasible for (P4)~\cite{cheng2024networked}, and they achieve the same objective value as the relaxed problem (P4.SDR), ensuring that the relaxation is tight. This iterative process guarantees convergence to a high-quality locally optimal solution.

\subsection{Receive Filter Optimization}

In this section, we focus on optimizing the receive filter $\mathbf{u}_{j,k}[n]$ under the given beamforming matrices $\mathbf{W}_{m,k}^{c}[n]$ and $\mathbf{W}_{m,k}^{r}[n]$, as well as the fixed UAV trajectory $\mathbf{q}_{k}[n]$. 

To begin with, the sensing SINR $\gamma_{k}^{r}[n]$ can be reformulated as
\begin{align}
    \gamma_{k}^{r}[n] = \sum_{j \in \mathcal{M}} \frac{\mathbf{u}_{j,k}^{H}[n] \mathbf{E}_{j,k}[n] \mathbf{u}_{j,k}[n]}{\mathbf{u}_{j,k}^{H}[n] \mathbf{F}_{j,k}[n] \mathbf{u}_{j,k}[n]},
\end{align}
where the matrices $\mathbf{E}_{j,k}[n]$ and $\mathbf{F}_{j,k}[n]$ are defined as follows

\begin{align}
    \mathbf{E}_{j,k}[n]={}& \sigma_{t}^{2}\mathbf{A}_{j,k}[n]\mathbf{X}_{j}[n]\mathbf{A}_{j,k}^{H}[n],\\
    \mathbf{F}_{j,k}[n]={}& \zeta_{j} \mathbf{H}_{\mathrm{SI},j}\mathbf{X}_{j}[n]\mathbf{H}_{\mathrm{SI},j}^{H} \nonumber\\
    &+\sum_{i\ne j,i\in \mathcal{M}}\alpha_{i,j} \mathbf{G}_{i,j}\mathbf{X}_{i}[n]\mathbf{G}_{i,j}^{H} +\sigma_{r}^{2}\mathbf{I}_{L} \nonumber\\
    &+\sigma_{t}^{2}\sum_{l\ne k,l\in\mathcal{K}}\mathbf{A}_{j,l}[n]\mathbf{X}_{j}[n]\mathbf{A}_{j,l}^{H}[n],
\end{align}
where $\mathbf{E}_{j,k}[n]$ represents the covariance of the desired echo signals for UAV $k$, and $\mathbf{F}_{j,k}[n]$ is the covariance matrix of the interference and noise. Both $\mathbf{E}_{j,k}[n]$ and $\mathbf{F}_{j,k}[n]$ are both semi-positive Hermitian matrices.

The optimization of the receive filter \( \mathbf{u}_{j,k}[n] \) can be formulated as follows
\begin{subequations}
\label{eq:problem_P5}
\begin{align}
(\text{P}5):\quad &\max_{\{\mathbf{u}_{j,k}[n]\}} \sum_{n \in \mathcal{N}} R[n] \nonumber\\
\text{s.t.}\quad &\sum_{j \in \mathcal{M}} \frac{\mathbf{u}_{j,k}^{H}[n] \mathbf{E}_{j,k}[n] \mathbf{u}_{j,k}[n]}{\mathbf{u}_{j,k}^{H}[n] \mathbf{F}_{j,k}[n] \mathbf{u}_{j,k}[n]} \geq \Gamma.
\end{align}
\end{subequations}

The expression of the sensing SINR \( \gamma_{k}^{r}[n] \) is in the form of a generalized Rayleigh quotient, where both the numerator and denominator are quadratic forms in \( \mathbf{u}_{j,k}[n] \). It is well known that the optimal solution to such a quotient is achieved when \( \mathbf{u}_{j,k}[n] \) is the dominant (i.e., principal) eigenvector of the matrix \( \mathbf{F}_{j,k}^{-1/2}[n] \mathbf{E}_{j,k}[n] \mathbf{F}_{j,k}^{-1/2}[n] \), and the corresponding optimal value of the sensing SINR equals the largest eigenvalue.

Therefore, to satisfy the sensing SINR constraint \( \gamma_k^r[n] \geq \Gamma \), it is sufficient to ensure that this maximum eigenvalue is no less than \( \Gamma \). Let \( \lambda_{j,k}^{\max}[n] \) denote the largest eigenvalue of \( \mathbf{F}_{j,k}^{-1/2}[n] \mathbf{E}_{j,k}[n] \mathbf{F}_{j,k}^{-1/2}[n] \). The problem \text{(P5)} can then be equivalently reformulated as
\begin{subequations}
\label{eq:problem_P6}
\begin{align}
(\text{P}6):\quad &\max_{\{\mathbf{u}_{j,k}[n]\}} \sum_{n \in \mathcal{N}} R[n] \nonumber\\
\text{s.t.}\quad &\sum_{j \in \mathcal{M}} \lambda_{j,k}^{\max}[n] \geq \Gamma.
\end{align}
\end{subequations}

The optimal receive filter \( \mathbf{u}_{j,k}^{*}[n] \) is thus obtained as the eigenvector associated with the largest eigenvalue of the matrix \( \mathbf{F}_{j,k}^{-1/2}[n] \mathbf{E}_{j,k}[n] \mathbf{F}_{j,k}^{-1/2}[n] \). This solution ensures that the sensing SINR is maximized under the given constraint and that \( \mathbf{u}_{j,k}[n] \) is aligned with the most favorable eigenmode of the effective sensing channel~\cite{liu2024cooperative}.

\subsection{UAV Trajectory Optimization}

In this section, our goal is to optimize the UAV trajectory design $\mathbf{q}_k[n]$, while keeping the beamforming vectors $\mathbf{W}_{m,k}^{c}[n]$ and $\mathbf{W}_{m,k}^{r}[n]$, as well as the receive filter $\mathbf{u}_{j,k}[n]$, fixed. The optimization problem is formally expressed as
\begin{align}
    (\text{P}7): & \max_{\{\mathbf{q}_k[n]\}} \sum_{n \in \mathcal{N}} R[n] \nonumber \\
    \text{s.t.} & \text{~\eqref{eq:problem1_b}, ~\eqref{eq:problem1_d}, ~\eqref{eq:problem1_e}, ~\eqref{eq:problem1_f}, ~\eqref{eq:problem1_g}.} \nonumber
\end{align}

Constraints~\eqref{eq:problem1_d} and~\eqref{eq:problem1_e} ensure that the UAVs adhere to the specified initial and final positions,~\eqref{eq:problem1_f} imposes a maximum velocity limit on UAV movement, and~\eqref{eq:problem1_g} enforces collision avoidance between UAVs. However, constraint~\eqref{eq:problem1_g} is nonconvex, making direct optimization challenging. The trajectory variable $\mathbf{q}_k[n]$ appears in $\theta_{m,k}[n]$, which directly impacts the achievable rate $R_k[n]$. Therefore, it is necessary to process the nonconvexity of~\eqref{eq:problem1_g} and reformulate $R_k[n]$ to facilitate optimization.

To handle this, we first apply the SCA method to approximate the non-convex constraint~\eqref{eq:problem1_g}. In the $f$-th iteration, the UAV trajectory variable $\mathbf{q}_k[n]$ is represented as $\mathbf{q}_k^{(f)}[n]$. Using the first-order Taylor expansion around $\mathbf{q}_k^{(f)}[n]$, the original constraint (20g) is approximated as:
\begin{align}
    -\Vert \mathbf{q}_j^{(f)}[n] - \mathbf{q}_i^{(f)}[n] \Vert^2 
    + 2 (\mathbf{q}_j^{(f)}[n] - \mathbf{q}_i^{(f)}[n])^T (\mathbf{q}_j[n] - \mathbf{q}_i[n]) \nonumber \\
    \geq D_{\min}^2 - (H_j - H_i)^2.
    \label{eq:q}
\end{align}

This reformulation transforms the original non-convex constraint into a convex one, making it easier to be solved by using convex optimization techniques.

Next, we reformulate the achievable communication rate $R_k[n]$ in the objective function for ease of derivation. To simplify the process, we adopt the following notations:
\begin{itemize}
    \item The entries in the $p$-th row and $q$-th column of the beamforming matrices $\mathbf{W}_{m,i}^{c}[n]$ and $\mathbf{W}_{m,i}^{r}[n]$ are denoted as $[\mathbf{W}_{m,i}^{c}[n]]_{p,q}$ and $[\mathbf{W}_{m,i}^{r}[n]]_{p,q}$, respectively.
    \item The absolute values of these entries are represented as $\vert[\mathbf{W}_{m,i}^{c}[n]]_{p,q}\vert$ and $\vert[\mathbf{W}_{m,i}^{r}[n]]_{p,q}\vert$.
    \item The phases of these entries are denoted as $\theta_{p,q}^{c}$ and $\theta_{p,q}^{r}$, respectively.
\end{itemize}

Based on these notations, the achievable rate $R_k[n]$ can be expressed in a simplified form, facilitating further derivation and optimization. This reformulation allows for precise handling of the non-linearities in $R_k[n]$ and serves as a basis for iterative optimization using convex techniques. The following relationship can be established as
\begin{align}
    R_{k}[n]
    = {}& \log_{2}(\sum_{m \in \mathcal{M}}\sum_{i \in \mathcal{K}_{cs}}\eta_{m,i,k}[n]+\sum_{m \in \mathcal{M}}\sum_{i \in \mathcal{K}}\mu_{m,i,k}[n] \nonumber\\
    &+\sum_{m \in \mathcal{M}}\frac{\sigma^2}{\kappa}(\Vert \mathbf{q}_{k}[n]-\mathbf{v}_{m}\Vert ^2+H_{k}^2)) \nonumber\\
    {}&- \log_{2}(\sum_{m \in \mathcal{M}}\sum_{i \in \mathcal{K}_{cs} \backslash k}\eta_{m,i,k}[n]+\sum_{m \in \mathcal{M}}\sum_{i \in \mathcal{K}}\mu_{m,i,k}[n] \nonumber\\
    &+\sum_{m \in \mathcal{M}}\frac{\sigma^2}{\kappa}(\Vert \mathbf{q}_{k}[n]-\mathbf{v}_{m}\Vert ^2+H_{k}^2)),
    \label{eq:Rkn}
\end{align}
where $\eta_{m,i,k}[n]$ and $\mu_{m,i,k}[n]$ are expressed as
\begin{align}
    &\eta_{m,i,k}[n]= \sum_{z=1}^{L}[\mathbf{W}_{m,i}^{c}[n]]_{z,z}+2\sum_{p=1}^{L}\sum_{q=p+1}^{L}\vert[\mathbf{W}_{m,i}^{c}[n]]_{p,q}\vert \nonumber\\
    {}& \quad \times \cos(\theta_{p,q}^{c}+2\pi\frac{d}{\lambda}(q-p)\frac{H_{k}}{\sqrt{\Vert \mathbf{q}_{k}[n]-\mathbf{v}_{m}\Vert ^2+H_{k}^2}}),
\end{align}
and
\begin{align}
    &\mu_{m,i,k}[n]= \sum_{z=1}^{L}[\mathbf{W}_{m,i}^{r}[n]]_{z,z}+2\sum_{p=1}^{L}\sum_{q=p+1}^{L}\vert[\mathbf{W}_{m,i}^{c}[n]]_{p,q}\vert \nonumber\\
    {}& \quad \times \cos(\theta_{p,q}^{r}+2\pi\frac{d}{\lambda}(q-p)\frac{H_{k}}{\sqrt{\Vert \mathbf{q}_{k}[n]-\mathbf{v}_{m}\Vert ^2+H_{k}^2}}).
\end{align}

Subsequently, we approximate~\eqref{eq:Rkn} by using the first-order Taylor expansion in iteration $f$ as
\begin{align}
    R_{k}[n] \approx {}& \tilde{R}_{k}^{(f)}[n] \nonumber\\
    ={}& c_{k}^{(f)}[n]+\mathbf{d}_{k}^{(f)T}[n](\mathbf{q}_{k}[n]-\mathbf{q}_{k}^{(f)}[n])
    \label{eq:approximate}
\end{align}
where
\begin{align}
    c_{k}^{(f)}[n]= {}& \log_{2}(\sum_{m \in \mathcal{M}}\sum_{i \in \mathcal{K}_{cs}}\eta_{m,i,k}^{(f)}[n]+\sum_{m \in \mathcal{M}}\sum_{i \in \mathcal{K}}\mu_{m,i,k}^{(f)}[n] \nonumber\\
    &+\sum_{m \in \mathcal{M}}\frac{\sigma^2}{\kappa}(\Vert \mathbf{q}_{k}^{(f)}[n]-\mathbf{v}_{m}\Vert ^2+H_{k}^2)) \nonumber\\
    {}&- \log_{2}(\sum_{m \in \mathcal{M}}\sum_{i \in \mathcal{K}_{cs} \backslash k}\eta_{m,i,k}^{(f)}[n] \nonumber\\
    &+\sum_{m \in \mathcal{M}}\sum_{i \in \mathcal{K}}\mu_{m,i,k}^{(f)}[n] \nonumber\\
    &+\sum_{m \in \mathcal{M}}\frac{\sigma^2}{\kappa}(\Vert \mathbf{q}_{k}^{(f)}[n]-\mathbf{v}_{m}\Vert ^2+H_{k}^2)),
\end{align}
and
\begin{align}
    \mathbf{d}_{k}^{(f)}[n]={}&\frac{\log_{2}e}{\phi_{k}[n]}(\sum_{m \in \mathcal{M}}\sum_{i \in \mathcal{K}_{cs}}\gamma_{m,i,k}^{(f)}[n] \nonumber\\
    {}&+\sum_{m \in \mathcal{M}}\sum_{i \in \mathcal{K}}\omega_{m,i,k}^{(f)}[n](\mathbf{q}_{k}^{(f)}[n]-\mathbf{v}_{m})) \nonumber\\
    {}&-\frac{\log_{2}e}{\psi_{k}[n]}(\sum_{m \in \mathcal{M}}\sum_{i \in \mathcal{K}_{cs} \backslash k}\gamma_{m,i,k}^{(f)}[n] \nonumber\\
    {}&+\sum_{m \in \mathcal{M}}\sum_{i \in \mathcal{K}}\omega_{m,i,k}^{(f)}[n](\mathbf{q}_{k}^{(f)}[n]-\mathbf{v}_{m})),
\end{align}
where $\gamma_{m,i,k}^{(f)}[n]$, $\omega_{m,i,k}^{(f)}[n]$, $\phi_{k}[n]$ and $\psi_{k}[n]$ are defined respectively as follows
\begin{align}
    &\gamma_{m,i,k}^{(f)}[n] = \sum_{p=1}^{L}\sum_{q=p+1}^{L}4\pi\vert[\mathbf{W}_{m,i}^{c}[n]]_{p,q}\vert \nonumber\\
    {}&\quad \times \sin(\theta_{p,q}^{c}+2\pi\frac{d}{\lambda}(q-p)\frac{H_{k}}{\sqrt{\Vert \mathbf{q}_{k}^{(f)}[n]-\mathbf{v}_{m}\Vert ^2+H_{k}^2}}) \nonumber\\
    {}&\quad \times \frac{dH_{k}(q-p)}{\lambda(\Vert \mathbf{q}_{k}^{(f)}[n]-\mathbf{v}_{m}\Vert ^2+H_{k}^2)^\frac{3}{2}},
\end{align}
\begin{align}
    &\omega_{m,i,k}^{(f)}[n] = \sum_{p=1}^{L}\sum_{q=p+1}^{L}4\pi\vert[\mathbf{W}_{m,i}^{r}[n]]_{p,q}\vert \nonumber\\
    {}&\quad \times \sin(\theta_{p,q}^{c}+2\pi\frac{d}{\lambda}(q-p)\frac{H_{k}}{\sqrt{\Vert \mathbf{q}_{k}^{(f)}[n]-\mathbf{v}_{m}\Vert ^2+H_{k}^2}}) \nonumber\\
    {}&\quad \times \frac{dH_{k}(q-p)}{\lambda(\Vert \mathbf{q}_{k}^{(f)}[n]-\mathbf{v}_{m}\Vert ^2+H_{k}^2)^\frac{3}{2}},
\end{align}
\begin{align}
    \phi_{k}[n]={}&\sum_{m \in \mathcal{M}}\sum_{i \in \mathcal{K}_{cs}}\eta_{m,i,k}^{(f)}[n] +\sum_{m \in \mathcal{M}}\sum_{i \in \mathcal{K}}\mu_{m,i,k}^{(f)}[n] \nonumber\\
    &+\sum_{m \in \mathcal{M}}\frac{\sigma^2}{\kappa}(\Vert \mathbf{q}_{k}^{(f)}[n]-\mathbf{v}_{m}\Vert ^2+H_{k}^2),\\
    \psi_{k}[n]={}&\sum_{m \in \mathcal{M}}\sum_{i \in \mathcal{K}_{cs} \backslash k}\eta_{m,i,k}^{(f)}[n] +\sum_{m \in \mathcal{M}}\sum_{i \in \mathcal{K}}\mu_{m,i,k}^{(f)}[n] \nonumber\\
    &+\sum_{m \in \mathcal{M}}\frac{\sigma^2}{\kappa}(\Vert \mathbf{q}_{k}^{(f)}[n]-\mathbf{v}_{m}\Vert ^2+H_{k}^2).
\end{align}

Similarly, we treat the constraint~\eqref{eq:problem1_b} in a form related to $\mathbf{q}_{k}[n]$ and guarantee that the constraint is convex.

Denote that $\tilde{\mathbf{A}}_{m,k}[n] = \mathbf{A}_{m,k}^{H}[n] \mathbf{u}_{m,k}[n] \mathbf{u}_{m,k}^{H}[n] \mathbf{A}_{m,k}[n]$, so as to $\tilde{\mathbf{H}}_{\text{SI},j}$ and $\tilde{\mathbf{G}}_{i,j}$ then the equation~\eqref{eq:SINR_re} can be reformulated in equation~\eqref{eq:SINR_RE}. 

\begin{figure*}
    \begin{equation}
        \gamma_{k}^r[n] = \frac{\sigma_{t}^{2} \displaystyle\sum_{j \in \mathcal{M}} \text{tr}(\tilde{\mathbf{A}}_{j,k}[n]\mathbf{X}_{j}[n])}
        {\displaystyle\sum_{j \in \mathcal{M}} [\text{tr}(\tilde{\mathbf{H}}_{\text{SI},j}\mathbf{X}_{j}[n])+\text{tr}(\tilde{\mathbf{G}}_{i,j}\mathbf{X}_{j}[n])+\sigma_{r}^{2}\mathbf{I}_{L} + \sigma_{t}^{2}\sum_{l \ne k, l \in \mathcal{K}}\text{tr}(\tilde{\mathbf{A}}_{j,k}[n]\mathbf{X}_{j}[n])]}.
        \label{eq:SINR_RE}
    \end{equation}
\end{figure*}

Let $\mathbf{u}_{m,k}[n] = [\alpha_{1}, \alpha_{2}, ..., \alpha_{L}]^{T}$, $\tilde{\mathbf{a}}_{m,k}[n] = \mathbf{A}^{H}_{m,k}[n] \cdot \mathbf{u}_{m,k}[n] = \mathbf{A}_{m,k}[n] \cdot \mathbf{u}_{m,k}[n]$, then we have $[\tilde{\mathbf{a}}_{m,k}[n]]_{i} = \sum_{q = 1}^{L}\alpha_{q} \cdot e^{j2\pi \frac{d}{\lambda}(i-q)\cos\theta_{m,k}[n]}$, $1 \le k \le L$. Since $\tilde{\mathbf{A}}_{m,k}[n] = \tilde{\mathbf{a}}_{m,k}[n] \cdot \tilde{\mathbf{a}}_{m,k}^{H}[n]$, $[\tilde{\mathbf{A}}_{m,k}[n]]_{p,q}$ can be expressed as

\begin{align}
    [\tilde{\mathbf{A}}_{m,k}[n]]_{p,q} = &{}(\sum^{L}_{s=1}\alpha_{s}e^{j2\pi \frac{d}{\lambda}(p - s)\cos\theta_{m,k}[n]})  \nonumber \\&\cdot(\sum^{L}_{t=1}\alpha_{t}^{*}e^{-j2\pi \frac{d}{\lambda}(q - t)\cos\theta_{m,k}[n]}) \nonumber \\
    = &{}\sum_{s=1}^{L}\sum_{t=1}^{L}\alpha_{m}\alpha_{n}^{*}e^{j2\pi \frac{d}{\lambda}[p - q - (s - t)]\cos\theta_{m,k}[n]}.
\end{align}

Next, we approximate $\text{tr}(\tilde{\mathbf{A}}_{j,k}[n]\mathbf{X}_{j}[n])$ by using the first-order Taylor expansion in iteration $f$ as

\begin{align}
    \text{tr}(\tilde{\mathbf{A}}_{j,k}[n]\mathbf{X}_{j}[n]) = h^{(f)}_{j,k}[n] + \boldsymbol{i}^{(f)}_{j,k}[n](\mathbf{q}_{k}[n]-\mathbf{q}_{k}^{(f)}[n]),
    \label{eq:trA}
\end{align}

where

\begin{align}
    h^{(f)}_{j,k}[n] = & \sum^{L}_{s=1} \sum^{L}_{t=1} \alpha_{s}\alpha^{*}_{t} \nonumber\\
    & \cdot[\sum^{L}_{z = 1}[\mathbf{X}_{j}[n]]_{z,z} + 2 \sum^{L}_{p = 1}\sum^{L}_{q = p + 1}\vert [\mathbf{X}_{j}[n]]_{z,z}\vert \nonumber \\
    & \times\cos(\theta^{x}_{p,q} + 2 \pi \frac{d}{\lambda}\frac{(p - q + t - s)H_{k}}{\sqrt{\Vert\mathbf{q}^{(f)}_{k}[n]-\mathbf{v}_{m}\Vert^{2} + H^{2}_{k}}})],
\end{align}

and

\begin{align}
    \boldsymbol{i}^{(f)}_{j,k}[n] =  & \sum^{L}_{s=1} \sum^{L}_{t=1} \alpha_{s}\alpha^{*}_{t} \cdot [ \sum^{L}_{p = 1} \sum^{L}_{q = p + 1} 4 \pi \vert \mathbf{X}_{j,k}[n] \nonumber \\
    & \cdot \sin(\theta^{x}_{p,q} + \frac{2 \pi d (p - q + t - s)H_{k}}{\sqrt{\Vert\mathbf{q}^{(f)}_{k}[n]-\mathbf{v}_{m}\Vert^{2} + H^{2}_{k}}})
    \nonumber \\
    &\times \frac{d(p - q + t - s)H_{k}}{\lambda (\Vert \mathbf{q}^{(f)}_{k}[n] - \mathbf{v}_{m} \Vert^{2} + H^{2}_{k})^{\frac{3}{2}}}],
\end{align}
in which $\theta^{x}_{p,q}$ represents the phase of the entries.

By bringing the formula~\eqref{eq:trA} into~\eqref{eq:SINR_RE}, the constraint~\eqref{eq:problem1_b} can be transformed into the form of the constraint~\eqref{eq:ge_gamma}. It is clear that the inequality constraint is linear with respect to $\mathbf{q}_{k}[n]$, and therefore the constraint is convex.

\begin{figure*}
    \begin{align}
        &\sigma_{t}^{2} \sum_{j \in \mathcal{M}} [h^{(f)}_{j,k}[n] + \boldsymbol{i}^{(f)}_{j,k}[n](\mathbf{q}_{k}[n]-\mathbf{q}_{k}^{(f)}[n])] \nonumber\\
        \ge & \Gamma \cdot \sum_{j \in \mathcal{M}} \{\text{tr}(\tilde{\mathbf{H}}_{\text{SI},j}\mathbf{X}_{j}[n])+\text{tr}(\tilde{\mathbf{G}}_{i,j}\mathbf{X}_{j}[n])+\sigma_{r}^{2}\mathbf{I}_{L} + \sigma_{t}^{2}\sum_{l \ne k, l \in \mathcal{K}}[h^{(f)}_{j,k}[n] + \boldsymbol{i}^{(f)}_{j,k}[n](\mathbf{q}_{k}[n]-\mathbf{q}_{k}^{(f)}[n])]\} .
    \label{eq:ge_gamma}
    \end{align}
\end{figure*}

To address the non-convexity of the objective function in equation~\eqref{eq:Rkn}, we reformulate it into a linearized form, as presented in equation~\eqref{eq:approximate}. This reformulation simplifies the problem and makes it suitable for convex optimization techniques. However, to ensure the accuracy of the linear approximation at each iterative step, a mechanism called the trust region constraint~\cite{lyu2022joint} is introduced. The trust region constraint is mathematically expressed as
\begin{align}
    \|\mathbf{q}_{k}^{(f)}[n] - \mathbf{q}_{k}^{(f-1)}[n]\| \leq \varepsilon^{(f)}, \quad \forall k \in \mathcal{K}, \, n \in \mathcal{N},
    \label{eq:trust_region}
\end{align}
where $\varepsilon^{(f)}$ represents the radius of the trust region in the $f$-th iteration. This constraint ensures that the trajectory update between successive iterations remains within a bounded region, maintaining the validity of the first-order approximation.

By incorporating the trust region constraint along with the approximated forms of the objective and constraints, the original nonconvex problem (P7) is reformulated as the convex problem (P8) for the $f$-th iteration. This reformulated problem is expressed as:
\begin{align}
    (\text{P}8): & \max_{\{\mathbf{q}_k[n]\}} \sum_{n \in \mathcal{N}} \sum_{k \in \mathcal{K}} \omega_k \tilde{R}_k^{(f)}[n] \nonumber \\
    \text{s.t.} & \quad \text{~\eqref{eq:problem1_b}, ~\eqref{eq:problem1_e}, ~\eqref{eq:problem1_f}, ~\eqref{eq:q}, ~\eqref{eq:ge_gamma}, ~\eqref{eq:trust_region}.} \nonumber
\end{align}

This convex approximation problem can be efficiently solved using numerical tools such as CVX, a solver designed for convex optimization.

\begin{algorithm}[t]
\caption{Trajectory Optimization Algorithm for Problem (P8)}
\begin{algorithmic}[1]
\State \textbf{Initialize:} UAV trajectory $\{\hat{\mathbf{q}}^{(0)}[n]\}$, trust region $\psi^{(0)}$, outer iteration index $f = 1$
\Repeat
    \State Let $l = 1$, set $\{\mathbf{q}^{(l-1)}[n]\} = \{\hat{\mathbf{q}}^{(f-1)}[n]\}$
    \Repeat
        \State Solve problem (P8) under local point
        $\{\mathbf{q}^{(l-1)}[n], \mathbf{W}_k^{(f)}[n], \mathbf{R}_s^{(f)}[n]\}$
        to obtain $\{\mathbf{q}^{(l)*}[n]\}$
        \If {objective value of problem (P8) increases}
            \State $\{\mathbf{q}^{(l)}[n]\} = \{\mathbf{q}^{(l)*}[n]\}$, $l \gets l + 1$
        \Else
            \State Update trust region: $\psi^{(l)} = \psi^{(l)} / 2$
        \EndIf
    \Until{$\psi^{(l)} < \hat{\varepsilon}$}
    \State Update UAV trajectory: $\{\hat{\mathbf{q}}^{(f)}[n]\} = \{\mathbf{q}^{(l)}[n]\}$, $f \gets f + 1$
\Until{the objective value converges within tolerance $\bar{\varepsilon}$}
\end{algorithmic}
\label{alg:sca_traj}
\end{algorithm}

In summary, the overall algorithm for solving problem (P7) operates iteratively by solving a sequence of convex subproblems. Each subproblem refines the solution using the trust region and linearized approximations. Similarly, the algorithms for the subproblems (P3), (P5), and (P7) converge by either finding the optimal solution or by successive refinements of the approximations. As a result, the  the AO technique is guaranteed to converge to a locally optimal solution for the joint UAV trajectory design and beamforming optimization problem.

\section{Numerical Results}
This section presents numerical results to illustrate the performance of our proposed design. In the simulation, we set up $K_{cs}=2$ ISAC UAVs, $K_{s}=2$ sensing UAVs , and $M = 4$ BSs near the UAV trajectory. The antenna spacing is set as $d = \frac{\lambda}{2}$, and each BS is equipped with $L = 8$ antennas. The starting positions of the two ISAC UAVs are $\mathbf{q}^{\mathrm{I}}_{1}$ = [50\,m, 150\,m] , $\mathbf{q}^{\mathrm{I}}_{2}$ = [50\,m, 450\,m], and the final positions of them are $\mathbf{q}^{\mathrm{F}}_{1}$ = [550\,m, 150\,m], $\mathbf{q}^{\mathrm{F}}_{2}$ = [550\,m, 450\,m]. The starting positions of the two sensing UAVs are $\mathbf{q}^{\mathrm{I}}_{3}$ = [50\,m, 250\,m], $\mathbf{q}^{\mathrm{I}}_{4}$ = [50\,m, 350\,m], and the corresponding final positions are $\mathbf{q}^{\mathrm{F}}_{3}$ = [550\,m, 250\,m], $\mathbf{q}^{\mathrm{F}}_{4}$ = [550\,m, 350\,m]. We set $H_{k}$ = 100\,m , devide $T$ into $N = 30$ time slots, and the maximum flight speed of each UAV is $V_{\max} = 20$\,m/s. The maximum transmit power of BSs is $P_{\max} = 10$\,W. Furthermore, the channel power gain at a reference distance of $d_{0} = 1$\,m is given as $-45$ dB, while the noise power at UAVs is $-100$\,dBW. The RCS of each UAV is modeled as a unit complex Gaussian variable with $\xi_{j,l} \sim \mathcal{CN}(0, 1)$, the self-interference coefficient $\zeta_{j,l}$ is set as -110\,dB, and the inter-base-station interference coefficient $\alpha_{j,l}$ is set to $-30$\,dB. All the weighting parameters $\omega_k$ are uniformly set to 1. For comparison, we set the following benchmark designs. 

\begin{figure*}[t]
    \centering
    \begin{subfigure}[b]{0.48 \textwidth}  
        \centering
        \includegraphics[width=\textwidth]{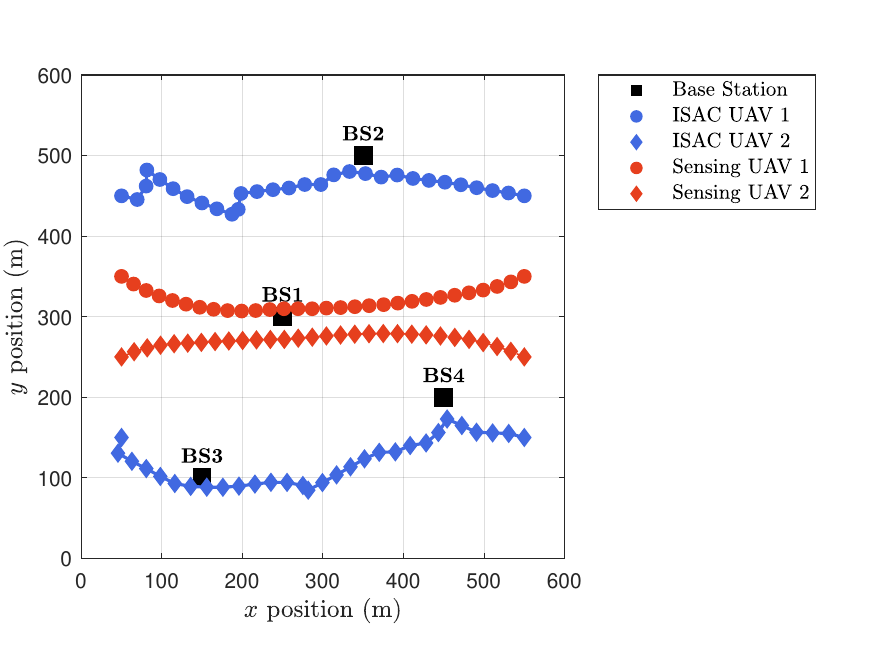}
        \caption{$\Gamma$ = -7\,dB, $H_{k}$ = 100\,m}
        \label{fig:image1}
    \end{subfigure}
    \hfill
    \begin{subfigure}[b]{0.48 \textwidth}
        \centering
        \includegraphics[width=\textwidth]{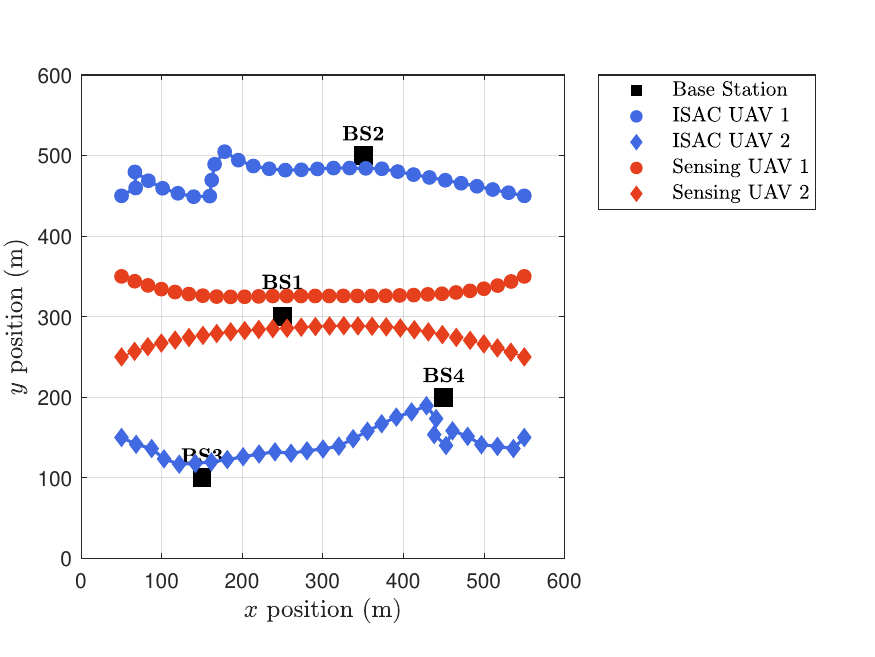}
        \caption{$\Gamma$ = -12\,dB, $H_{k}$ = 100\,m}
        \label{fig:image2}
    \end{subfigure}

    \vspace{0.5cm}

    \begin{subfigure}[b]{0.48 \textwidth}
        \centering
        \includegraphics[width=\textwidth]{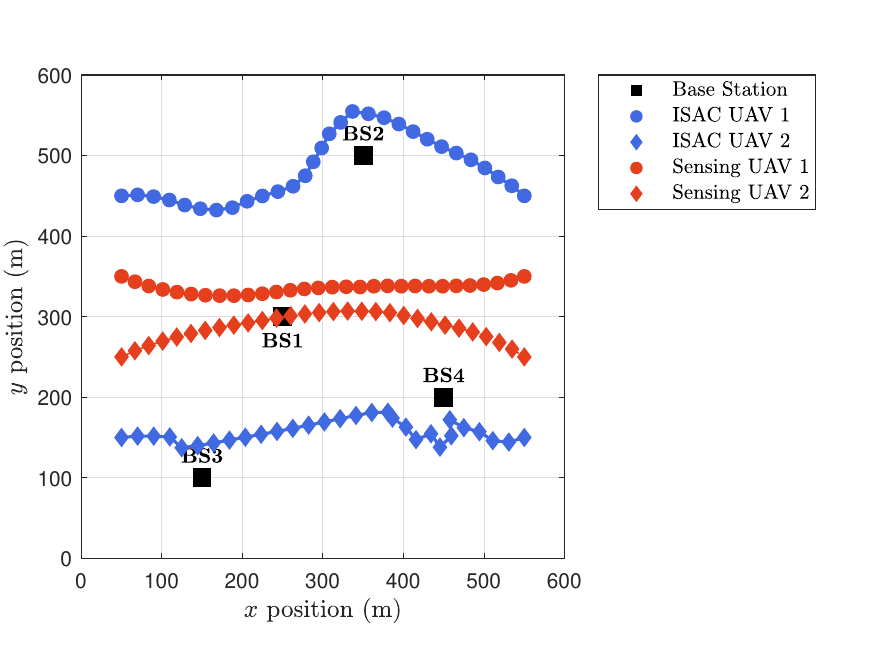}
        \caption{$\Gamma$ = -7\,dB, $H_{k}$ = 120\,m}
        \label{fig:image3}
    \end{subfigure}
    \hfill
    \begin{subfigure}[b]{0.48 \textwidth}
        \centering
        \includegraphics[width=\textwidth]{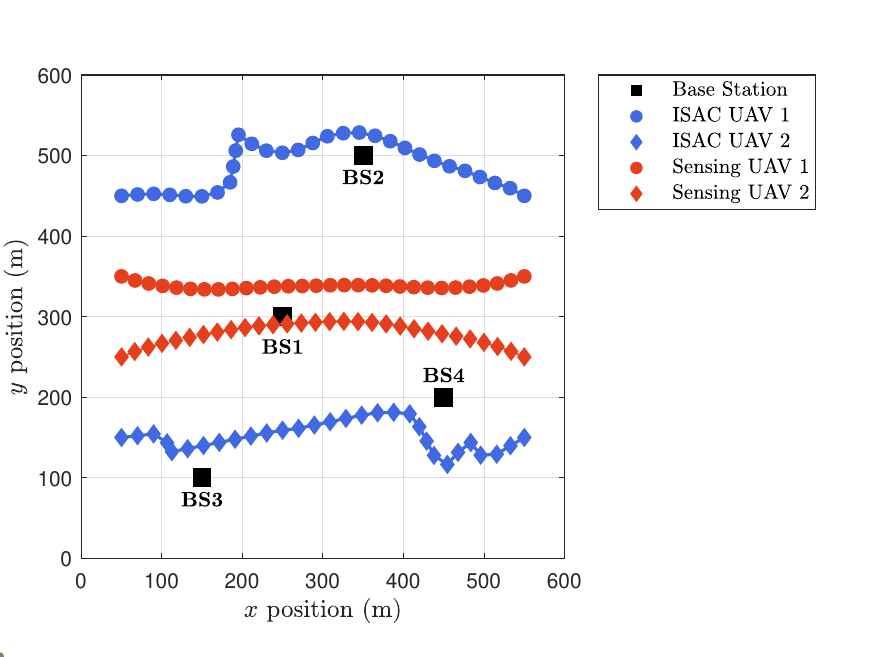}
        \caption{$\Gamma$ = -12\,dB, $H_{k}$ = 100\,m}
        \label{fig:image4}
    \end{subfigure}

    \caption{Optimal UAV trajectories under different altitudes and sensing thresholds.}
    \label{fig:2x2grid}
\end{figure*}

\textbf{Optimized beamforming with a uniform straight-line trajectory}: 
In this scenario, the trajectories of the four UAVs are predetermined and not subject to optimization. Each UAV moves along a straight-line path from its initial position to its final destination without deviation. The motion follows a constant velocity given by \( V_{k} = \frac{1}{N} \Vert \mathbf{q}^{\mathrm{I}}_{k} - \mathbf{q}^{\mathrm{F}}_{k} \Vert \), where \( \mathbf{q}^{\mathrm{I}}_{k} \) and \( \mathbf{q}^{\mathrm{F}}_{k} \) represent the initial and final positions of UAV \( k \), respectively. Given that the movement of the UAVs is constrained to a uniform linear trajectory, the beamforming matrices $\{\mathbf{W}^{c}_{m,i}[n]\}$  and $\{\mathbf{W}^{r}_{m,i}[n]\}$ are optimized to enhance communication performance by solving problem (P4), ensuring effective signal transmission throughout the entire flight duration.

\textbf{Optimized beamforming and straight-line trajectory with optimized speed}: 
This setting extends the previous benchmark by allowing the UAVs to adjust their speeds along the straight-line trajectories instead of maintaining a uniform velocity. While the UAVs still follow straight paths, their speeds are dynamically optimized to maximize the overall sum communication rate. This flexibility in speed control allows the UAVs to linger longer in areas where improved connectivity can be achieved, leading to enhanced communication performance. To achieve this, we jointly optimize the beamforming matrices \( \{\mathbf{W}^{c}_{m,i}[n]\} \), \( \{ \mathbf{W}^{r}_{m,i}[n]\} \), as well as the UAV positions \( \{\mathbf{q}_{k}[n]\} \), by solving problem (P2) under the additional constraint that the trajectories remain linear but with variable speeds.

\textbf{Optimized trajectory, not beamforming}: 
In this approach, we focus on optimizing the UAV trajectories while maintaining a simple equal power allocation beamforming strategy. Instead of dynamically adapting the beamforming weights, we assume equal power distribution to simplify the beamforming process. Specifically, we set the beamforming matrices as \( \mathbf{W}^{c}_{m,i}[n] = \frac{p^{c}_{j,l}[n]}{L}\mathbf{I}_{L} \) and \( \mathbf{W}^{r}_{m,i}[n] = 
\frac{p^{r}_{j,l}[n]}{L}\mathbf{I}_{L} \), where \( p^{c}_{j,l}[n] \) and \( p^{r}_{j,l}[n] \) represent the power allocated to the communication and sensing signals, respectively. These power variables are constrained by the total power budget at each BS, given by \( \sum_{i \in \mathcal{K}_{cs}} p^{c}_{j,l}[n] + \sum_{i \in \mathcal{K}} p^{r}_{j,l}[n] \leq P_{\max} \). Given the fixed beamforming strategy, the UAV trajectories \( \mathbf{q}_{k}[n] \) are then optimized by solving problem (P8) with the pre-defined beamforming matrices \( \mathbf{W}^{c}_{m,i}[n] \) and \( \mathbf{W}^{r}_{m,i}[n] \). This approach ensures that the UAVs adjust their flight paths to achieve optimal communication performance despite not having adaptive beamforming.

Through analyzing the optimized UAV trajectories shown in Fig. \ref{fig:2x2grid}, we can explore the impact of different sensing SINR thresholds ($\Gamma$) and UAV flight altitudes ($H_k$) on trajectory optimization.

It is obvious that the sensing SINR threshold has a significant impact on UAV trajectories. This can be observed by comparing Fig. \ref{fig:image1} and Fig. \ref{fig:image2} at $H_k = 100$\,m, and Fig. \ref{fig:image3} and Fig. \ref{fig:image4} at $H_k = 120$\,m. In both cases, when the sensing threshold increases from $\Gamma = -12$\,dB to $\Gamma = -7$\,dB, the UAV trajectories become more concentrated and closer to the BSs. This is because a higher sensing SINR requirement forces UAVs to adjust their flight paths to ensure that the BSs can receive stronger echo signals, thereby enhancing sensing performance. In contrast, a lower threshold allows the UAVs greater flexibility in their trajectories, enabling a more balanced optimization between communication throughput and sensing quality. A high sensing threshold thus leads to sensing-driven trajectory planning, while a lower threshold grants more degrees of freedom to enhance communication performance.

The flight altitude of UAVs is also a critical factor in trajectory optimization. This can be seen by comparing Fig. \ref{fig:image1} and Fig. \ref{fig:image3} for $\Gamma = -7$\,dB, and Fig. \ref{fig:image2} and Fig. \ref{fig:image4} for $\Gamma = -12$\,dB. When the altitude increases from $H_k = 100$\,m to $H_k = 120$\,m, noticeable changes occur in the trajectory distribution. Higher altitudes increase the propagation distance of sensing echoes, leading to more severe path loss and lower sensing SINR at the BSs. Communication links are similarly affected, resulting in degraded transmission quality. In contrast, lower-altitude UAVs benefit from better channel conditions, allowing more flexible and effective trajectory optimization. These observations indicate that UAV altitude should be carefully configured in accordance with specific mission requirements, as joint optimization of trajectory and altitude plays a crucial role in balancing communication and sensing objectives.

\begin{figure}[h]
    \centering
    \includegraphics[width=0.8\linewidth]{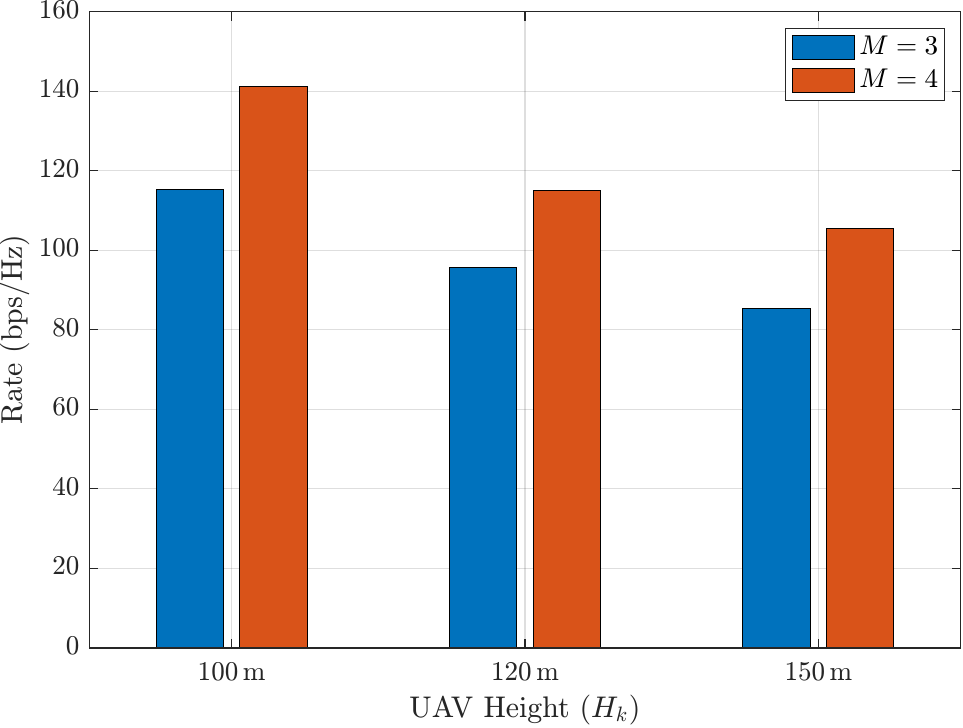}
    \caption{Weighted sum rate versus UAV flight altitude $H_k$ under different numbers of BSs ($\Gamma = -12$\,dB).}
    \label{fig:height_bs_rate}
\end{figure}

Fig.~\ref{fig:height_bs_rate} illustrates the impact of UAV altitude on the weighted sum rate under different BS deployment densities, with the sensing SINR threshold fixed at $\Gamma = -12$~dB. The results show that the communication rate declines as the UAV altitude increases, primarily due to the aggravated path loss between the UAVs and ground BSs, which deteriorates the communication channel quality. Furthermore, a higher number of BSs significantly improves the system performance. In particular, the three-base-station configuration is obtained by excluding one BS from the original four-base-station layout. Across all tested altitudes, the four-base-station deployment consistently outperforms its counterpart, owing to the increased spatial diversity and enhanced beamforming degrees of freedom. Although additional BSs may introduce more potential interference, the availability of higher spatial resolution enables more effective interference management and signal enhancement through coordinated beamforming. These findings highlight the necessity of jointly optimizing UAV altitude and BS deployment to achieve superior ISAC performance.

\begin{figure}[h]
    \centering
    \includegraphics[width=0.8\linewidth]{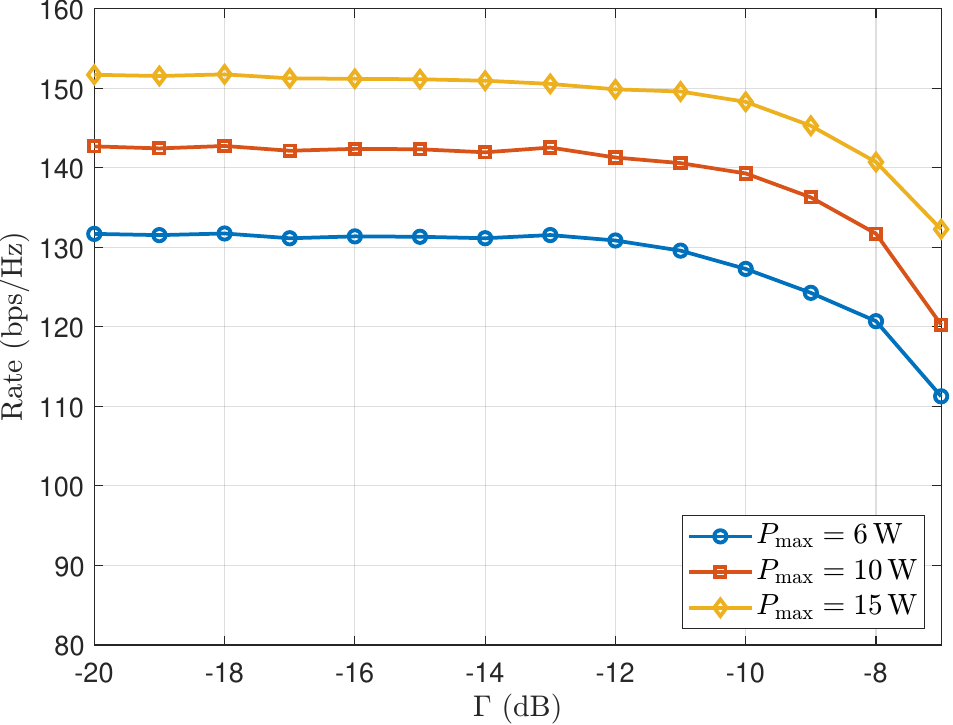}
    \caption{Weighted sum rate versus sensing SINR threshold $\Gamma$ under different BS transmit power limits $P_\text{max}$.}
    \label{fig:power_rate_gamma}
\end{figure}

Fig.~\ref{fig:power_rate_gamma} presents the relationship between the weighted sum rate and the sensing SINR threshold $\Gamma$ under different maximum transmit power constraints of each BS ($P_{\max} = 6$\,W, 10\,W, and 15\,W). As $\Gamma$ increases, the system is required to meet stricter sensing performance, which in turn limits the degrees of freedom available for communication. Consequently, the overall communication rate shows a clear decreasing trend with the increase of the sensing SINR threshold. Moreover, increasing the maximum transmit power of BSs can effectively alleviate the trade-off between sensing and communication. Higher power budgets provide more resources to simultaneously support both tasks, resulting in better communication performance even under stringent sensing requirements. These results suggest that appropriately increasing the BS power can enhance system robustness and improve the performance ceiling of ISAC systems.

\begin{figure}[h]
    \centering
    \includegraphics[width=0.8\linewidth]{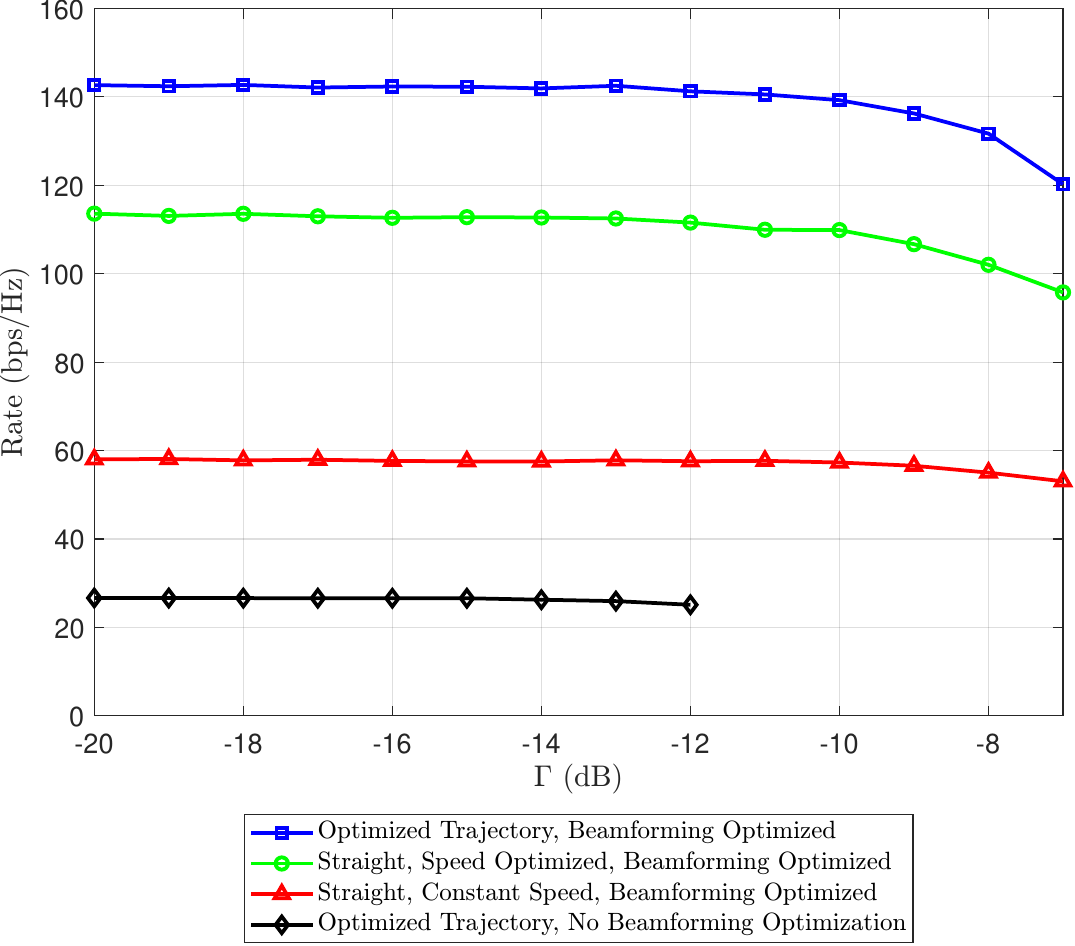}
    \caption{Weighted sum rate versus sensing SINR threshold $\Gamma$ under different trajectory and beamforming optimization strategies.}
    \label{fig:benchmark}
\end{figure}

Fig.~\ref{fig:benchmark} illustrates the impact of different trajectory and beamforming optimization schemes on the system communication rate under varying sensing SINR thresholds $\Gamma$. As $\Gamma$ increases, the communication rate of all schemes declines to different extents, depending on the specific optimization strategy. This is because the BSs are required to allocate more transmit power and spatial resources to satisfy the sensing constraints, which limits the degrees of freedom available for communication. Nonetheless, the joint trajectory and beamforming optimization scheme consistently achieves the highest communication rate across all $\Gamma$ values, demonstrating its superior performance and adaptability to changing requirements.

The scheme with a fixed straight-line trajectory but optimized speed performs reasonably well across the entire range of $\Gamma$ values, yet still falls short of the fully optimized trajectory scheme. This suggests that adjusting UAV speed alone provides limited performance gains, and that spatial trajectory design plays a more essential role in aligning UAVs with favorable channel conditions to improve communication.

In contrast, the scheme with fixed trajectory and constant speed but optimized beamforming yields significantly lower communication rates, indicating that beamforming alone is insufficient to handle increasing sensing requirements without the aid of trajectory adaptation. Its performance remains mostly flat and suboptimal across the range of thresholds.

Finally, the scheme with optimized trajectory but no beamforming optimization achieves the lowest communication rates and becomes infeasible at higher $\Gamma$ values, resulting in a truncated curve. This highlights the importance of beamforming in maintaining sensing feasibility, as trajectory design alone cannot ensure acceptable performance under strict constraints. Overall, Fig.~\ref{fig:benchmark} shows that joint trajectory and beamforming optimization is crucial for maximizing communication performance and ensuring robustness under varying sensing demands.

\section{Conclusion}

This paper investigated a cooperative UAV-enabled ISAC system, where UAV trajectories and transmit beamforming were jointly optimized to maximize communication performance while satisfying sensing quality constraints. We formulated an optimization problem to maximize the weighted sum rate under sensing SINR constraints, while accounting for UAV mobility and power limitations. To solve the resulting non-convex problem, an alternating optimization framework was proposed, incorporating SDR and SCA techniques. Simulation results showed that the joint optimization approach outperforms benchmark schemes, delivering higher communication throughput and robust sensing performance. The results also underscored the value of trajectory adaptation under varying sensing SINR requirements and UAV altitudes.

\bibliographystyle{ieeetr}
\bibliography{references}

\begin{thebibliography}{10}

\bibitem{zhang2024cooperative}
Y.~Zhang, H.~Shan, Y.~Zhou, Z.~Shi, L.~Sheng, and Y.~Liu, ``Cooperative beamforming design for anti-{UAV} {ISAC} systems,'' {\em IEEE Trans. Wireless Commun.}, vol.~24, pp.~2249--2264, Mar. 2025.

\bibitem{meng2023uav}
K.~Meng, Q.~Wu, J.~Xu, W.~Chen, Z.~Feng, R.~Schober, and A.~L. Swindlehurst, ``{UAV}-enabled integrated sensing and communication: Opportunities and challenges,'' {\em IEEE Wirel. Commun.}, vol.~31, pp.~97--104, Apr. 2024.

\bibitem{cheng2024networked}
G.~Cheng, X.~Song, Z.~Lyu, and J.~Xu, ``Networked {ISAC} for low-altitude economy: Transmit beamforming and {UAV} trajectory design,'' in {\em Proc. IEEE/CIC International Conference on Communications in China (ICCC)}, pp.~78--83, Aug. 2024.

\bibitem{long2020joint}
H.~Long, M.~Chen, Z.~Yang, Z.~Li, B.~Wang, X.~Yun, and M.~Shikh-Bahaei, ``Joint trajectory and passive beamforming design for secure {UAV} networks with {RIS},'' in {\em Proc. IEEE Globecom Workshops (GC Wkshps)}, pp.~1--6, Dec. 2020.

\bibitem{luo2022joint}
Y.~Xu, Y.~Li, J.~A. Zhang, M.~D. Renzo, and T.~Q.~S. Quek, ``Joint beamforming for {RIS}-assisted integrated sensing and communication systems,'' {\em IEEE Trans. Commun.}, vol.~72, pp.~2232--2246, Apr. 2024.

\bibitem{pan2023cooperative}
Y.~Pan, R.~Li, X.~Da, H.~Hu, M.~Zhang, D.~Zhai, K.~Cumanan, and O.~A. Dobre, ``Cooperative trajectory planning and resource allocation for {UAV}-enabled integrated sensing and communication systems,'' {\em IEEE Trans. Veh. Technol.}, vol.~73, pp.~6502--6516, May 2024.

\bibitem{jing2024isac}
X.~Jing, F.~Liu, C.~Masouros, and Y.~Zeng, ``{ISAC} from the sky: {UAV} trajectory design for joint communication and target localization,'' {\em IEEE Trans. Wireless Commun.}, vol.~23, pp.~12857--12872, Oct. 2024.

\bibitem{lyu2022joint}
Z.~Lyu, G.~Zhu, and J.~Xu, ``Joint maneuver and beamforming design for {UAV}-enabled integrated sensing and communication,'' {\em IEEE Trans. Wireless Commun.}, vol.~22, pp.~2424--2440, Apr. 2023.

\bibitem{khalili2024efficient}
A.~Khalili, A.~Rezaei, D.~Xu, F.~Dressler, and R.~Schober, ``Efficient {UAV} hovering, resource allocation, and trajectory design for {ISAC} with limited backhaul capacity,'' {\em IEEE Trans. Wireless Commun.}, vol.~23, pp.~17635--17650, Nov. 2024.

\bibitem{wang2024isac}
Y.~Wang, K.~Zu, L.~Xiang, Q.~Zhang, Z.~Feng, J.~Hu, and K.~Yang, ``Isac enabled cooperative detection for cellular-connected {UAV} network,'' {\em IEEE Trans. Wireless Commun.}, vol.~24, pp.~1541--1554, Feb. 2025.

\bibitem{chai2024precoding}
R.~Chai, X.~Cui, R.~Sun, D.~Zhao, and Q.~Chen, ``Precoding and trajectory design for {UAV}-assisted integrated communication and sensing systems,'' {\em IEEE Trans. Veh. Technol.}, vol.~73, pp.~13151--13163, Sep. 2024.

\bibitem{meng2022uav}
K.~Meng, Q.~Wu, S.~Ma, W.~Chen, and T.~Q.~S. Quek, ``{UAV} trajectory and beamforming optimization for integrated periodic sensing and communication,'' {\em IEEE Wireless Commun. Lett.}, vol.~11, pp.~1211--1215, Jun. 2022.

\bibitem{liu2024cooperative}
S.~Liu, R.~Liu, Z.~Lu, M.~Li, and Q.~Liu, ``Cooperative cell-free {ISAC} networks: Joint {BS} mode selection and beamforming design,'' in {\em Proc. IEEE Wireless Commun. Netw. Conf. (WCNC)}, pp.~1--6, Apr. 2024.

\bibitem{zhang2024joint}
R.~Zhang, Y.~Zhang, R.~Tang, H.~Zhao, Q.~Xiao, and C.~Wang, ``A joint {UAV} trajectory, user association, and beamforming design strategy for multi-{UAV}-assisted {ISAC} systems,'' {\em IEEE Internet Things J.}, vol.~11, pp.~29360--29374, Sep. 2024.

\bibitem{gao2024trajectory}
Q.~Gao, R.~Zhong, and Y.~Liu, ``Trajectory and beamforming optimization in {UAV}-enabled {ISAC} system,'' in {\em Proc. IEEE Global Commun. Conf. (GLOBECOM)}, pp.~1527--1532, Dec. 2024.

\bibitem{pang2024dynamic}
X.~Pang, S.~Guo, J.~Tang, N.~Zhao, and N.~Al-Dhahir, ``Dynamic {ISAC} beamforming design for {UAV}-enabled vehicular networks,'' {\em IEEE Trans. Wireless Commun.}, vol.~23, pp.~16852--16864, Nov. 2024.

\bibitem{pei2024joint}
F.~Pei, L.~Xiang, and A.~Klein, ``Joint optimization of beamforming and 3d array-steering for {UAV}-aided {ISAC},'' in {\em Proc. IEEE Int. Conf. Commun. (ICC)}, pp.~1249--1254, Jun. 2024.

\bibitem{zhou2024temporal}
S.~Zhou, H.~Yang, L.~Xiang, and K.~Yang, ``Temporal-assisted beamforming and trajectory prediction in sensing-enabled {UAV} communications,'' {\em IEEE Trans. Commun.}, pp.~1--1, Dec. 2024.

\bibitem{he2023full}
Z.~He, W.~Xu, H.~Shen, D.~W.~K. Ng, Y.~C. Eldar, and X.~You, ``Full-duplex communication for {ISAC}: Joint beamforming and power optimization,'' {\em IEEE J. Sel. Areas Commun.}, vol.~41, pp.~2920--2936, Sep. 2023.

\bibitem{wu2018joint}
Q.~Wu, Y.~Zeng, and R.~Zhang, ``Joint trajectory and communication design for multi-{UAV} enabled wireless networks,'' {\em IEEE Trans. Wireless Commun.}, vol.~17, pp.~2109--2121, Mar. 2018.

\bibitem{yuan2022joint}
X.~Yuan, H.~Jiang, Y.~Hu, and A.~Schmeink, ``Joint analog beamforming and trajectory planning for energy-efficient {UAV}-enabled nonlinear wireless power transfer,'' {\em IEEE J. Sel. Areas Commun.}, vol.~40, pp.~2914--2929, Oct. 2022.

\bibitem{lin2023deep}
H.~Lin, Z.~Zhang, L.~Wei, Z.~Zhou, and T.~Zheng, ``A deep reinforcement learning based {UAV} trajectory planning method for integrated sensing and communications networks,'' in {\em Proc. IEEE Veh. Technol. Conf. (VTC)}, pp.~1--6, Oct. 2023.

\bibitem{pan2020multicell}
C.~Pan, H.~Ren, K.~Wang, W.~Xu, M.~Elkashlan, A.~Nallanathan, and L.~Hanzo, ``Multicell {MIMO} communications relying on intelligent reflecting surfaces,'' {\em IEEE Trans. Wireless Commun.}, vol.~19, pp.~5218--5233, Aug. 2020.

\bibitem{moon2024joint}
S.~Moon, H.~Liu, and I.~Hwang, ``Joint beamforming for {RIS}-assisted integrated sensing and secure communication in {UAV} networks,'' {\em J. Commun. Netw.}, vol.~26, pp.~502--508, Oct. 2024.

\bibitem{pang2021irs}
X.~Pang, N.~Zhao, J.~Tang, C.~Wu, D.~Niyato, and K.-K. Wong, ``{IRS}-assisted secure {UAV} transmission via joint trajectory and beamforming design,'' {\em IEEE Trans. Commun.}, vol.~70, pp.~1140--1152, Feb. 2022.

\bibitem{deng2024joint}
D.~Deng, W.~Zhou, X.~Li, D.~B. da~Costa, D.~W.~K. Ng, and A.~Nallanathan, ``Joint beamforming and {UAV} trajectory optimization for covert communications in {ISAC} networks,'' {\em IEEE Trans. Wireless Commun.}, vol.~24, pp.~1016--1030, Feb. 2025.

\bibitem{sankar2023beamforming}
R.~S.~P. Sankar, S.~P. Chepuri, and Y.~C. Eldar, ``Beamforming in integrated sensing and communication systems with reconfigurable intelligent surfaces,'' {\em IEEE Trans. Wireless Commun.}, vol.~23, pp.~4017--4031, May 2024.

\bibitem{li2023joint}
S.~Li, H.~Du, D.~Zhang, and K.~Li, ``Joint {UAV} trajectory and beamforming designs for {RIS}-assisted {MIMO} system,'' {\em IEEE Trans. Veh. Technol.}, vol.~73, pp.~5378--5392, Apr. 2024.

\bibitem{ge2020joint}
L.~Ge, P.~Dong, H.~Zhang, J.-B. Wang, and X.~You, ``Joint beamforming and trajectory optimization for intelligent reflecting surfaces-assisted {UAV} communications,'' {\em IEEE Access}, vol.~8, pp.~78702--78712, Apr. 2020.

\bibitem{xiu2024improving}
Y.~Xiu, W.~Lyu, P.~L. Yeoh, Y.~Ai, Y.~Li, and N.~Wei, ``Improving physical-layer security in {ISAC}-{AAV} system: Beamforming and trajectory optimization,'' {\em IEEE Trans. Veh. Technol.}, vol.~74, pp.~3503--3508, Feb. 2025.

\bibitem{zhang2023sensing}
X.~Zhang, M.~Peng, and C.~Liu, ``Sensing-assisted beamforming and trajectory design for {UAV}-enabled networks,'' {\em IEEE Trans. Veh. Technol.}, vol.~73, pp.~3804--3819, Mar. 2024.

\end{thebibliography}

\end{document}